\documentclass[sigconf,nonacm,pbalance]{acmart}
\usepackage{placeins}
\providecommand{\tagstructbegin}[1]{}
\providecommand{\tagstructend}{}
\providecommand{\tagpdfsetup}[1]{}
\makeatletter
\@ifundefined{KV@Gin@alt}{\define@key{Gin}{alt}{}}{}
\@ifundefined{KV@Gin@artifact}{\define@key{Gin}{artifact}[true]{}}{}
\makeatother

\setcopyright{none}
\acmDOI{}
\acmISBN{}
\hypersetup{pdfauthor={Zihong He, Hai-Ning Liang, Chen Liang},pdflang={en-US},pdfdisplaydoctitle=true}
\newcommand{\modelrowseparator}{\addlinespace[2pt]}
\newcommand{\historyrowseparator}{\addlinespace[2pt]}

\begin{document}
\title{Tap-to-Adapt: Aligning Response Initiation Timing in Interactive Speech Agents through Corrective Taps}

\author{Zihong He}
\affiliation{%
  \institution{The Hong Kong University of Science and Technology (Guangzhou)}
  \city{Guangzhou}
  \country{China}}
\email{zhe154@connect.hkust-gz.edu.cn}

\author{Hai-Ning Liang}
\authornote{Corresponding authors.}
\affiliation{%
  \institution{The Hong Kong University of Science and Technology (Guangzhou)}
  \city{Guangzhou}
  \country{China}}
\affiliation{%
  \institution{The Hong Kong University of Science and Technology}
  \city{Hong Kong}
  \country{China}}
\email{hainingliang@hkust-gz.edu.cn}

\author{Chen Liang}
\authornotemark[1]
\affiliation{%
  \institution{The Hong Kong University of Science and Technology (Guangzhou)}
  \city{Guangzhou}
  \country{China}}
\email{chenliang2@hkust-gz.edu.cn}
\renewcommand{\shortauthors}{He et al.}

\begin{abstract}
Aligning speech agents' response initiation with users' expectations remains challenging. We introduce Tap-to-Adapt, a framework for interactive speech systems in which manual activation and interruption provide both immediate response control and online supervision for personalized initiation timing. To learn from sparse corrections and recent feedback, we introduce \emph{Balanced Proximal Replay for History-conditioned Timing} (BPRHT) as its response initiation timing control mechanism, combining frozen speech representations, class-balanced replay with proximal regularization, and recent correction history. We evaluate acoustic backbones for BPRHT on recorded speech and compare update rules and recent-history configurations in simulation. The selected replay and history mechanisms lower simulated correction rates relative to their respective baselines. To reduce new users' cold-start correction burden, we evaluate shared initialization in a two-stage user study. Stage~1 established shared initialization with 20 participants, locking the average of earlier participants' final heads upon parameter and opening-correction plateaus. Stage~2 assigned 48 new participants equally to online adaptation with or without shared initialization, or frozen shared initialization. Online adaptation without shared initialization had more opening corrections per minute than with it (rate ratio 1.69, 95\% confidence interval 1.15--2.49). Final-block correction rates were similar across conditions. Online updates accompanied reported speaking and waiting adjustments. Declining corrections under frozen parameters also suggest user accommodation. Code and 1,000 labeled pilot recordings are available at: \mbox{\url{https://github.com/ZihongHe/tap_to_adapt}}.
\end{abstract}

\ccsdesc[500]{Human-centered computing~User studies}
\ccsdesc[300]{Human-centered computing~Systems and tools for interaction design}
\ccsdesc[300]{Computing methodologies~Online learning settings}
\keywords{human-agent interaction, speech agents, response initiation timing, online learning, turn-taking, human-in-the-loop learning}

\begin{teaserfigure}
 \tagstructbegin{tag=Figure,alt={Users activate a silent agent when a response is expected, supplying positive samples, or interrupt an unwanted response, supplying negative samples. Both feedback paths enter Tap-to-Adapt to learn when to respond and when to wait.}}
 \includegraphics[width=\textwidth,artifact]{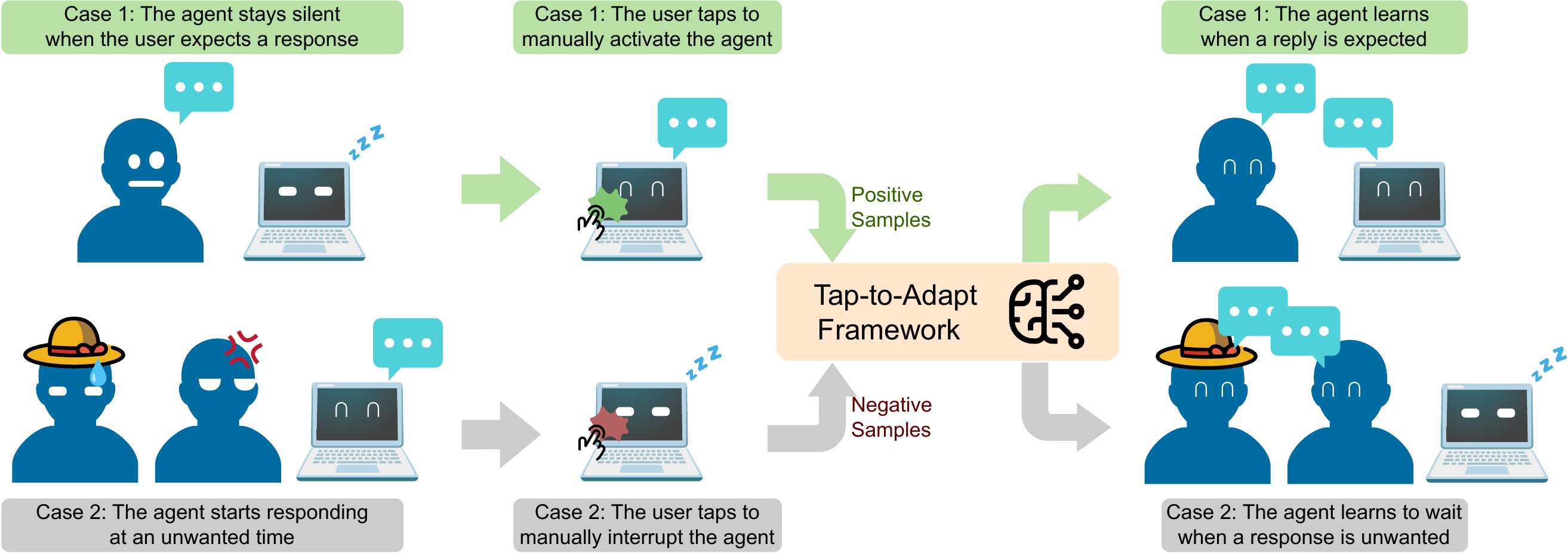}
 \caption{Overview of corrective interaction in the Tap-to-Adapt framework. Manual activation requests a response and supplies a positive initiation label, while manual interruption stops the current response and supplies a negative label for its initiation decision. These corrections guide subsequent response timing.}
 \Description{Two rows show complementary corrections. In the upper row, an agent remains silent when a reply is expected. The user taps to activate it, producing a positive training sample. In the lower row, the agent speaks at an unwanted time. An interruption tap produces a negative sample. Both feed the Tap-to-Adapt framework, whose response initiation timing control mechanism learns when to respond and when to wait.}
 \label{fig:teaser}
 \tagstructend
\end{teaserfigure}
\maketitle
\hypersetup{pdfauthor={Zihong He, Hai-Ning Liang, Chen Liang},pdfsubject={Response initiation timing, corrective taps, shared initialization, and co-adaptation in interactive speech agents}}

\section{Introduction}

Large language model (LLM)-enhanced interactive speech agents are extending natural-language interaction to mobile devices~\cite{deng2024mobile,liu2025llm}, wearables~\cite{mishrasing,ren2025toward}, smart homes~\cite{rivkin2024aiot,lin2025voicetalk}, in-vehicle systems~\cite{furusawa2024demonstration,furusawa2025vehicle}, and care for older adults~\cite{yang2024talk2care,liu2025toward}. Beyond generating contextually appropriate content, these systems must determine whether a response is needed and when to begin it. Delayed responses prolong waiting, whereas premature responses can interrupt an unfinished utterance. Response initiation timing therefore directly affects the flow and naturalness of turn-taking~\cite{skantze2021turn,walker2024natural,liu2025toward,ethiraj2025toward}.

Turn-initiation modeling and speech wake-up systems provide two foundations for response triggering. Turn-initiation models use acoustic and contextual cues to predict future speech activity, turn transitions, or utterance onsets. Their learning targets primarily derive from participants' observed speech events~\cite{skantze2017towards,ekstedt2022voice,li2022can,wagner2025survey}. Such supervision only partially represents when a user expects an agent to respond. Similar pauses or acoustic cues may call for continued waiting or a reply, depending on the user and context. Speech wake-up research detects particular words or phrases used to address a system~\cite{kepuska2009novel,kepuska2011wake,kepuska2017improving}. Explicit calling signals offer limited coverage of timing preferences during ongoing multi-turn conversations without a spoken invocation. However, the learning targets of these approaches differ from users' judgments of whether a particular response is timely, and mechanisms for continually calibrating individual preferences from timing corrections remain limited.

Interactive machine learning gives users a direct role in shaping model behavior through feedback~\cite{amershi2014power,knox2009interactively}. In speech interaction, pauses, repeated utterances, and overlapping speech can signal difficulty, with meanings that depend on context. A corrective tap provides an explicit request to start or stop a response and associates that request with a particular system decision.

To align response initiation with users' expectations during conversation, we introduce Tap-to-Adapt, a framework for interactive speech systems that couples immediate response control with online timing learning from corrective taps. It organizes response initiation, response generation, and long-term memory around this interaction loop. A \emph{manual activation} while the system is waiting supplies positive initiation supervision. A \emph{manual interruption} immediately terminates the current response and supplies negative supervision associated with \emph{that response's initiation decision}. Each tap is bound to the system state and the response decision it corrects, turning immediate interaction control into a learnable timing signal. This loop calibrates when the agent begins responding. The user-operated interruption control handles stopping after a response has begun. We use \emph{response timing} as shorthand for this initiation-timing objective throughout the paper.

To implement personalized timing learning within the framework, we introduce \emph{Balanced Proximal Replay for History-conditioned Timing} (BPRHT) as the response initiation timing control mechanism in our implementation. It combines multi-window representations from frozen DistilHuBERT with a user-specific logistic regression head, uses class-balanced replay with proximal regularization for online updates, and conditions response probability and the decision threshold on corrections and acoustic memory from the preceding 60 seconds. Section~\ref{method} details this mechanism within the design and implementation of the Tap-to-Adapt framework.

Section~\ref{val_exp_section} assesses the implemented timing loop and selects BPRHT's configuration through recorded-speech and simulation evaluations. On 1,000 labeled recordings, we compare acoustic backbones using F1, area under the precision--recall curve (AUPRC), Brier score, expected calibration error (ECE), and 95th-percentile (P95) inference latency. DistilHuBERT achieved the highest mean final F1 and AUPRC among tested candidates while meeting the inference-time budget. Simulated interactions on fixed LLM+TTS speech streams then compare online update rules and recent-history configurations. Class-balanced replay with proximal regularization yielded lower correction rates than gated and stabilized replay, and incorporating recent correction history yielded a lower mean correction rate than the no-history baseline. These comparisons guide BPRHT's configuration in the Tap-to-Adapt framework.

Building on this technical assessment, Section~\ref{usa_exp} examines the Tap-to-Adapt framework in live speech interaction through a two-stage user study. It investigates whether \emph{shared initialization} pretrained from earlier users reduces new users' correction burden and whether users and the system adapt to each other during use. The adaptation question draws on evidence of users' linguistic alignment in human-computer dialogue~\cite{branigan2010linguistic} and work on mutual adaptation between user behavior and agent policies~\cite{nikolaidis2017human}. \emph{Opening correction burden} denotes corrections in the first six-minute block.
\begin{itemize}
 \item \begin{minipage}[t]{\linewidth}\textbf{RQ1. Shared initialization and cold start.} How does shared initialization affect new users' cold-start correction burden in response timing alignment, and how does this effect change as interaction progresses?\end{minipage}
 \item \begin{minipage}[t]{\linewidth}\textbf{RQ2. User and system adaptation.} As interaction progresses, do users and the system exhibit mutual adaptation around the system's response initiation timing?\end{minipage}
\end{itemize}

Stage~1 recruits \emph{source participants} sequentially and uses a prespecified rule to assess saturation of the equal-weight average of their final decision heads. The rule was satisfied after 20 completions. The average of the first 18 participants' final heads became the fixed shared initialization for Stage~2. Stage~2 assigns 48 new users equally to \emph{cold-start online adaptation}, \emph{shared-initialization online adaptation}, and \emph{frozen after shared initialization}. Comparing the two adapting conditions addresses shared initialization's opening and subsequent effects for RQ1. For RQ2, we examine reported speaking and waiting adjustments and logged feedback-driven updates, and compare correction burden over time between shared-initialization adaptation and frozen shared initialization. Section~\ref{discussion} interprets the findings and discusses their implications, limitations, and future research.

Our contributions follow.
\begin{enumerate}
 \item Tap-to-Adapt, a framework for interactive speech systems that links manual activation and interruption to their associated initiation decisions, enabling immediate response control and online adaptation of response timing.
 \item BPRHT, which combines frozen speech representations, class-balanced replay with proximal regularization, and recent correction history under a predict-then-update protocol.
 \item A two-stage user study comprising shared-head pretraining and saturation assessment with 20 participants in Stage~1 and a three-condition evaluation with 48 new users in Stage~2. Shared-initialization online adaptation had lower opening correction burden than cold-start online adaptation. Final-block rates were similar, ranging from 0.65\% to 0.72\%. Online update records, frozen-condition declines, and reported strategy changes characterize system updating and user accommodation.

\end{enumerate}

\section{Related Work}

We review turn-initiation modeling, speech wake-up systems, and online speech learning from user feedback to establish how these research areas inform the design of the Tap-to-Adapt framework.
\subsection{Speech Turn-Initiation Modeling}
\label{related_work_speech_turn_init}

Aligning an interactive speech agent's turn onset with the user's expectations is an important goal in spoken interaction. Turn-initiation modeling predicts the onset of speech activity within the broader study of turn-taking, which also includes turn holding, turn switching, and backchannels~\cite{skantze2017towards,ekstedt2022voice,li2022can,wagner2025survey}. Skantze~\cite{skantze2017towards} proposed a continuous model using long short-term memory recurrent neural networks (LSTM-RNNs) to estimate each interlocutor's probability of speaking in future windows, frame by frame. Its pause-based evaluation selected cases in which only one person continued within the following second and predicted the next speaker, distinguishing turn holding from switching. The model also distinguished backchannels from longer turns at speech onset.

Ekstedt and Skantze~\cite{ekstedt2022voice} introduced Voice Activity Projection (VAP), using future voice activity as a self-supervised target. Their model combines representations from contrastive predictive coding (CPC), current and historical voice activity, and a causal Transformer to capture dependencies among future speech events. They evaluated turn switching and backchannel prediction in two-party dialogue. To address the limitations of pause-triggered prediction for rapid transitions and overlapping speech, Li et al.~\cite{li2022can} represented initiation as continuous time until the next utterance. Their model combines wav2vec 1.0 speech features, GPT-2 encodings of incremental transcripts, a recurrent network, and a Gaussian-mixture output to represent possible initiation times and uncertainty. Evaluation on Switchboard compared predicted and observed lead times.

These approaches establish a basis for continuous timing prediction using observed speech activity and utterance onsets. Personalizing response timing also calls for evidence of when each user expects the agent to speak. The same pause, intonation, or context may imply different preferred response times for different users.

Full-duplex models integrate response timing with streaming generation. Moshi jointly models user and agent speech streams, allowing silence, overlap, and speech onset to emerge during generation~\cite{defossez2024moshi}. MiniCPM-o 4.5 predicts listen or speak control tokens from time-aligned inputs before generating content~\cite{cui2026minicpm}.
~\citet{hsiao2026aspirin} optimized a projected binary policy over speech and silence using rule-based rewards for overlap and response latency.
~\citet{kim2026duplexgen} calibrated turn-taking actions for dialogue synthesis using human annotations for specific scenarios.
The Tap-to-Adapt framework builds on continuous timing prediction by using corrective taps to identify the timing decisions a user wants to change. In our implementation, BPRHT learns from these taps to adapt subsequent response timing to that user online.

\subsection{Speech Wake-Up Systems}

Speech wake-up activates a system through specific vocal signals~\cite{kepuska2009novel,kepuska2011wake,kepuska2017improving}. Applications include wearables~\cite{schmalfuss2024customized}, home voice assistants~\cite{ferro2024implementation}, and speech interfaces for older adults~\cite{yan2024understanding}. Early approaches commonly used hidden Markov models (HMMs), Gaussian mixture models (GMMs), or GMM-HMM hybrids. These parameterized models can operate with limited data, but robustness to speaker variation and noise remains a constraint~\cite{vroomen1992robust,ghaffarzadegan2017deep,schroder2016performance}.

Neural speech modeling offers richer acoustic representations and activation detection through waveform modeling, attention, convolutional and convolution-attention encoders, and large-scale speech-recognition pretraining~\cite{oord2016wavenet,shan2018attention,li2019jasper,gulati2020conformer,radford2023robust}. These methods address related but distinct tasks, including speech generation, recognition, and activation detection. For wake-word detection, Jose et al.~\cite{jose2020accurate} used convolutional neural networks (CNNs) to identify both the word and its onset and offset. Sahai et al.~\cite{sahai2023dual} proposed dual-attention neural transducers with separate attention paths for wake words and general vocabulary. Dynamic path selection improved wake-word recognition while reducing computation.

Most wake-up systems use fixed wake words as their primary trigger~\cite{kepuska2009novel,kepuska2011wake,kepuska2017improving,jose2020accurate}. Device-directed speech detection extends activation beyond keywords by determining whether speech is addressed to the system~\cite{mallidi2018device,norouzian2019exploring,dighe2024device}. In continuous dialogue, speech directed to the system may include pauses within an unfinished turn. The Tap-to-Adapt framework targets the timing at which a particular user expects a reply in keyword-free dialogue. In our implementation, BPRHT combines acoustic timing estimates with recent correction history to determine automatic activation. Manual activation and interruption provide direct supervision for response initiation timing.

\begin{figure}[t]
 \centering
 \includegraphics[width=\columnwidth,alt={The Waiting, Thinking, and Speaking layout shows activation and interruption around the user. The BPRHT mechanism receives speech and tap history. ASR, an LLM, and TTS form the response-generation path in the basic conversational setting. The LLM reads and updates long-term memory.}]{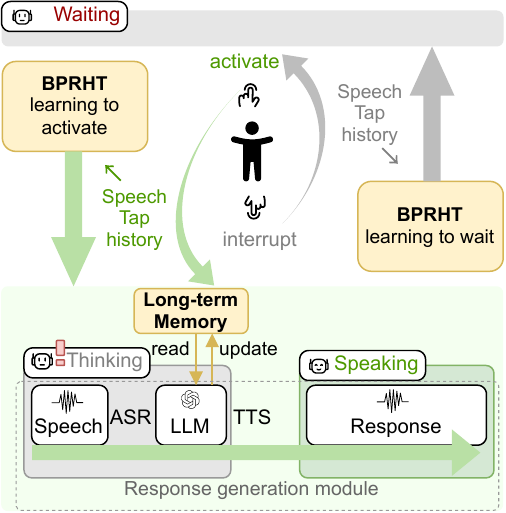}
 \caption{Functional modules of the Tap-to-Adapt framework and their evaluated implementation. Response generation and long-term memory provide the basic conversational setting, with their implementations fixed across timing conditions. Corrective taps combine immediate control with supervision of the associated initiation decision.}
 \Description{The Waiting, Thinking, and Speaking layout shows activation and interruption around the user. The BPRHT mechanism receives speech and tap history. ASR, an LLM, and TTS form the response-generation path in the basic conversational setting. The LLM reads and updates long-term memory.}
 \label{fig:system}
\end{figure}

\subsection{Online Speech Learning from User Feedback}

Interactive machine learning links feedback interface design, user understanding, and iterative model refinement~\cite{amershi2014power,dudley2018review}. TAMER learns from human evaluations of an agent's behavior~\cite{knox2009interactively}. Online speech learning from user feedback applies interactive machine learning to speech systems, using feedback from ongoing interactions to update their models.

Earlier work has used user feedback to optimize dialogue policies and models for speech processing online. Ga\v{s}i\'{c} et al.~\cite{gavsic2011line} collected binary satisfaction feedback during live conversations to optimize dialogue policies. Su et al.~\cite{su2015learning} predicted users' assessments of dialogue success from interaction features and used the learned reward for online reinforcement learning, reducing reliance on explicit ratings. Su et al.~\cite{su2016line} subsequently used active reward learning to select dialogues for which to request success feedback, supporting joint optimization of the reward model and policy.

In speech recognition, Zhou et al.~\cite{zhou2023gift} used users' manual transcript corrections on their devices to improve recognition of new and infrequent words through federated learning while controlling effects on overall recognition quality. He et al.~\cite{he2025interactive} accepted real-time verbal corrections in a streaming recognition and diarization pipeline, such as identifying the actual speaker of an earlier segment. Feedback immediately corrected speaker attribution, and online speaker enrollment reduced subsequent confusion. These studies demonstrate that corrections can provide continuing supervision for dialogue success, transcription, and speaker attribution.

Dialogue-level assessments, whether solicited or inferred from interaction, help optimize dialogue policies~\cite{gavsic2011line,su2015learning,su2016line}. These assessments can reflect several aspects of a conversation, making it difficult to tie the assessment to the timing of a particular response. Transcript and speaker-attribution corrections provide feedback about what was said and who said it~\cite{zhou2023gift,he2025interactive}. \citet{wu2025aligning} used preference pairs derived from spoken interactions to address content and timing errors through offline alignment of a full-duplex model.

The Tap-to-Adapt framework supports human-in-the-loop adaptation of response initiation. Users can correct an individual initiation decision during conversation, and BPRHT uses that feedback to guide subsequent timing decisions for the same user.
\section{Method}
\label{method}
In this section, we first describe the functional modules of the Tap-to-Adapt framework and their control and feedback loop. We then present BPRHT and its online learning procedure, followed by the evaluated implementation and its settings.

\begin{figure*}[t]
 \centering
 \includegraphics[width=\linewidth,trim=19.25bp 8.25bp 18.25bp 7.25bp,clip,alt={Frozen DistilHuBERT encodes up to 15 seconds of speech. Mean and standard-deviation pooling over seven windows produces fourteen 768-dimensional vectors. Concatenation and fixed normalization yield 10752 features. BPRHT combines a logistic regression head, logit calibration, and recent correction history for the response decision.}]{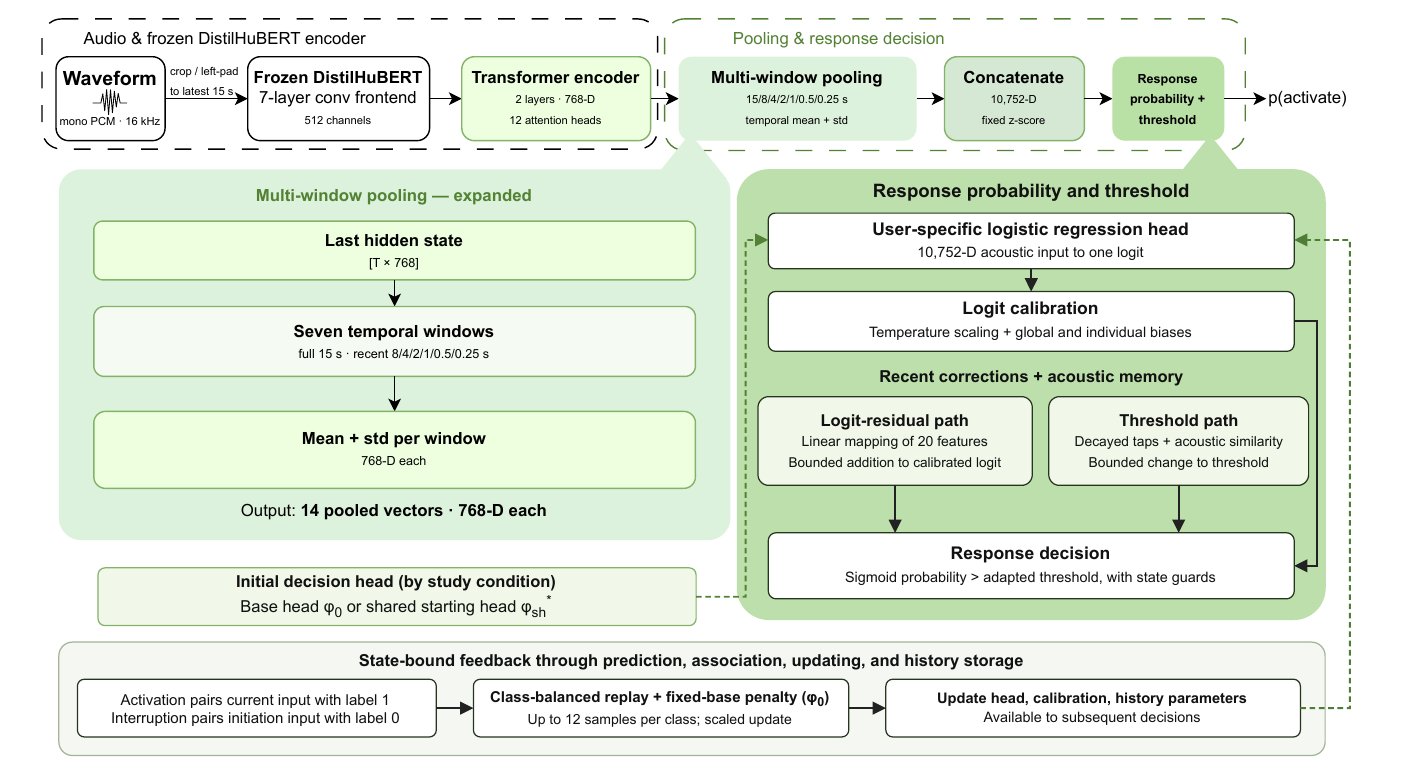}
 \caption{BPRHT, the response initiation timing control mechanism developed for the Tap-to-Adapt framework. Frozen DistilHuBERT extracts hidden states from up to 15 seconds of 16\,kHz PCM. Multi-window pooling computes temporal means and standard deviations at seven scales. Concatenation and fixed z-score normalization produce 10,752 features. The logistic regression head and logit calibration combine with history-dependent threshold and logit-residual adjustments to determine response initiation. Accepted corrections supervise their associated input before entering subsequent history. The encoder and normalization parameters remain fixed. In the user study, the initial user-specific head is the base head \(\phi_0\) or the locked shared starting head \(\phi_{\rm sh}^{*}\), depending on the condition (Section~\ref{extended_exp}).}
 \Description{A mono waveform is cropped or left-padded to 15 seconds, then passed through frozen DistilHuBERT's seven-layer convolutional frontend and two-layer, 768-dimensional Transformer encoder. The last hidden states are pooled over 15, 8, 4, 2, 1, 0.5, and 0.25 seconds. Each window produces a mean and standard-deviation vector, yielding fourteen 768-dimensional vectors. A logistic regression head produces one logit, and calibration and a history-dependent logit correction and threshold determine activation. A boxed initialization choice supplies either the base head or the shared starting head to the user-specific head. The encoder and fixed normalization are identified in the top row.}
 \label{fig:model}
\end{figure*}
\subsection{Framework Design}
\label{framework_overview}

The Tap-to-Adapt framework comprises three functional modules: response initiation, response generation, and long-term memory (Figure~\ref{fig:system}). The initiation module learns when to begin a response from ongoing speech and corrective taps. The generation and memory modules provide the basic conversational setting in which this initiation mechanism operates. Their implementations are kept fixed across the timing conditions evaluated here.
\paragraph{Response initiation.} We implement this module through \emph{Balanced Proximal Replay for History-conditioned Timing} (BPRHT), the response initiation timing control mechanism developed for this study. It reuses corrective feedback through class-balanced replay and constrains parameter drift with a proximal penalty toward a fixed base decision head. It combines acoustic timing estimates with recent correction history to adjust response probability and the decision threshold for automatic initiation.
This module models initiation as a continuous binary decision~\cite{inoue2024multilingual,castillo2025survey}. At time \(t\), it estimates whether responding is appropriate from speech \(a_{\leq t}\) and prior tap history \(\mathcal H_{<t}\). The system constructs labels from accepted control actions. Manual activation in Waiting pairs the current acoustic input with \(y=1\), indicating \emph{respond now}. Interruption in Thinking or Speaking pairs the stored input at that response's initiation with \(y=0\), indicating \emph{continue waiting}. Interruption also terminates generation or playback immediately. To preserve this association, the system cancels taps with mismatched state or response identifiers. An interruption supervises the earlier initiation decision. Stopping after initiation remains under user control. Section~\ref{study_interface} defines acceptance criteria.

\paragraph{Response generation.} This module prepares and delivers a spoken reply following an initiation decision and supports immediate cancellation by the user. Our evaluated implementation uses automatic speech recognition (ASR), an LLM, and text-to-speech (TTS) synthesis in a cascade. This function could also be implemented by an end-to-end speech model that generates audio without an intermediate text representation. In a full-duplex realization, initiation and generation could share a model, with generated speech, silence, or listen/speak control tokens governing whether a response begins.

\paragraph{Long-term memory.} This module supplies dialogue context for response generation and incorporates completed interactions into subsequent context. In the evaluated system, asynchronous LLM-generated summaries implement this role.

For integrated implementations, corrective taps would retain their immediate activation or interruption effects and supervise the corresponding initiation event, whether represented by an explicit decision or by model outputs.

\subsection{Response Initiation Timing Control}
\label{model_and_learning}

For lightweight online adaptation, BPRHT (defined in Section~\ref{framework_overview}) uses frozen DistilHuBERT~\cite{chang2022distilhubert} as the acoustic backbone and combines its representations with recent correction history, following the technical evaluation in Section~\ref{val_exp_section}. Figure~\ref{fig:model} shows the path from multi-window speech representations to response decisions. The response decision combines the calibrated output of a user-specific logistic regression head with history-dependent logit and threshold adjustments. Section~\ref{implementation_details} specifies the fixed implementation settings.

\subsubsection{Audio Input and Multi-Window Features}

To capture both recent acoustic changes and speech context, we extract features over multiple time windows. We crop or left-pad with zeros the latest 15\,s of mono 16\,kHz PCM for frozen DistilHuBERT. Its seven-layer convolutional frontend and two-layer, 768-dimensional Transformer encoder~\cite{vaswani2017attention} produce \(H_t=[h_1,\ldots,h_{J_t}]\in\mathbb{R}^{J_t\times768}\), where \(h_j\) is the encoded vector for frame \(j\) and \(J_t\) is the number of encoded frames. Over windows of 15, 8, 4, 2, 1, 0.5, and 0.25 seconds ending at \(t\), we compute elementwise temporal means and population standard deviations, adding a small positive constant \(\epsilon\) to variances before taking square roots for numerical stability. Concatenating these seven pairs yields the 10,752-dimensional vector \(z_t\).

To keep feature scales comparable across users and updates, we obtain \(\tilde z_t\) by subtracting the fixed development-set mean vector \(m\) and dividing elementwise by the fixed scale vector \(v\) plus \(\epsilon\). These z-score normalization~\cite{fawcett2021machine} parameters are shared across participants. Let \(\phi\) denote parameters of the response model and \(\phi_i=[w_i^\top,b_i]^\top\) participant \(i\)'s individual decision head. Its weight vector \(w_i\) and scalar bias \(b_i\) map \(\tilde z_t\) to the acoustic logit \(\ell_{\rm audio}(t)\). \(\top\) denotes transpose. Online learning updates each user's head and history-conditioning parameters, keeping DistilHuBERT and normalization fixed.

\subsubsection{History-conditioned Timing Decisions}

Recent corrections can help prevent recurring timing mismatches. BPRHT uses recent corrections to make subsequent decisions more conservative after interruptions and more responsive after activations. It summarizes accepted feedback from the 60\,s preceding the current decision in a 20-dimensional vector \(x_t\), then adjusts the response threshold and logit through parallel paths. To ground timing feedback in what the agent heard, we learn from accepted taps with usable audio for the corrected decision. Activation in Waiting uses current audio. Interruption in Thinking or Speaking uses the audio saved when the interrupted response was triggered. A decision-head update also requires a nonzero update scale \(s_t\). History includes accepted events that update calibration or history parameters during head warm-up. A \emph{speech prefix} is the audio available up to the decision instant, represented by its latest 15\,s. Features encode tap age, decay, action, interruption delay, and similarity or change relative to the stored input of an interrupted response, as detailed in Table~\ref{tab:history_features}. Subscripts \({\rm act}\) and \({\rm int}\) denote activation and interruption.
\begin{table*}[!t]
\centering
\caption{Recent correction-history features (20 dimensions). Ages \(a\) and \({\rm age}_e\) denote elapsed seconds since a tap and event \(e\), respectively. Missingness is 1 for missing and 0 for present, with missing numeric values set to zero. The fixed delay normalization scale is 3000\,ms. Normalized values may exceed 1. The fixed scale for feature differences is 8. The clip function bounds its argument to the specified interval, elementwise for vectors. \(\|\cdot\|_2\) denotes the Euclidean norm. If interruption features are available but the comparison prefix is missing, feature change defaults to 0.35.}
\label{tab:history_features}
\small
\begin{tabular}{@{}p{0.23\linewidth}p{0.72\linewidth}@{}}
\toprule
Feature family (dimensions) & Definition \\
\midrule
Time since tap (4) & Ages \(a_{\rm act},a_{\rm int}\) of the latest activation and interruption, each transformed as \(\log(1+a)/\log(1+60)\), with a missingness indicator. \\
Multi-scale decay (6) & Sum of \(\exp(-{\rm age}_e/\rho)\) over same-type events \(e\) in the window, separately for both actions, with time constants \(\rho=60,30\) seconds. The two 15-second channels are reserved for the quarter-window comparison and set to zero in the deployed configuration. \\
Latest action (2) & Direction of the latest action, with activation \(=+1\), interruption \(=-1\), no history \(=0\); plus an indicator for any event in the window, with present \(=1\) and absent \(=0\). \\
Interruption delay (2) & Latest available interval \(d_{\rm ms}\), in milliseconds, from activation notification to interruption, transformed as \(\log(1+d_{\rm ms})/\log(1+3000)\), with a missingness indicator. \\
Decayed feedback strength (2) & \(p_{\rm int}=\exp(-a_{\rm int}/25)\) and \(p_{\rm act}=\exp(-a_{\rm act}/40)\), with fixed time constants of 25 and 40 seconds. Set to zero if the event is missing or at least 60 seconds old. \\
Acoustic similarity (2) & Cosine similarity between features of the current prefix and the activation prefix of the latest interrupted response, mapped linearly to \([0,1]\), with a missingness indicator. \\
Feature change (2) & \({\rm clip}(\|\Delta\|_2/8,0,1)\), where \(\Delta\) is the feature difference between the interrupted response's activation prefix and the most recent prefix available at that interruption; plus a missingness indicator. \\
\bottomrule
\end{tabular}
\end{table*}

The threshold path changes the probability required for activation. Learned scalar gains \(k_{\rm int}\) and \(k_{\rm act}\) weight the decayed interruption and activation strengths \(p_{\rm int}\) and \(p_{\rm act}\) defined in Table~\ref{tab:history_features}. Acoustic similarity and feature change further modulate the interruption component. Subtraction, clipping, and a bias \(b_h\) yield a bounded threshold shift. The residual path applies a single linear mapping with weights \(w_h\) to \(x_t\), clips its output, and adds the resulting residual to the calibrated acoustic logit. The trainable parameters \(\phi_h=(k_{\rm int},k_{\rm act},b_h,w_h)\) comprise three scalars and this 20-dimensional vector.

The explicit threshold shift is
\[
\begin{aligned}
s_{\rm explicit}={\rm clip}\!\bigl(&k_{\rm int}p_{\rm int}\gamma_{\rm ac}\gamma_{\rm ab}\\
&-k_{\rm act}p_{\rm act},-\delta_{\rm exp}^{-},\delta_{\rm exp}^{+}\bigr).
\end{aligned}
\]
We retain some influence from interruption feedback across acoustic contexts through lower modulation bounds \(\alpha_{\rm ac}\) and \(\alpha_{\rm ab}\). Given acoustic features for the current input and for the start decision of the latest interrupted response, the similarity and feature-change multipliers are \(\gamma_{\rm ac}=\alpha_{\rm ac}+(1-\alpha_{\rm ac})s\) and \(\gamma_{\rm ab}=\alpha_{\rm ab}+(1-\alpha_{\rm ab})d\), respectively, where \(s\) is cosine similarity mapped to \([0,1]\) and \(d\) is the clipped feature-change magnitude defined in Table~\ref{tab:history_features}. If either vector is missing, the multipliers use fixed fallback values \(\bar\gamma_{\rm ac}\) and \(\bar\gamma_{\rm ab}\), respectively. Calibration adjusts the sharpness and offset of acoustic response probabilities using temperature \(T\), global bias \(b_g\), and individual bias \(b_i^{\rm cal}\). The two paths combine with the acoustic output as follows.
\begin{align}
\Delta\tau_t&={\rm clip}(s_{\rm explicit}+b_h,-\delta_\tau^{-},\delta_\tau^{+}),\\
\tau'_t&={\rm clip}(\tau_0+\Delta\tau_t,\tau_{\min},\tau_{\max}),\\
\ell_{\rm early}(t)&={\rm clip}(x_t^\top w_h,-L_h,L_h),\\
p_t&=\operatorname{sigmoid}\!\bigl(\ell_{\rm audio}(t)/T+b_g\notag\\
&\qquad+b_i^{\rm cal}+\ell_{\rm early}(t)\bigr).
\end{align}
Here \(\Delta\tau_t\) shifts the baseline threshold \(\tau_0\) to the adapted threshold \(\tau'_t\), \(\ell_{\rm early}(t)\) is the history-based logit residual, and the logistic sigmoid yields response probability \(p_t\).

Positive shift limits \((\delta_{\rm exp}^{-},\delta_{\rm exp}^{+})\) and \((\delta_\tau^{-},\delta_\tau^{+})\) bound threshold decreases and increases before and after adding \(b_h\), respectively. Larger upward limits allow more scope to delay responses after interruptions. Final bounds \((\tau_{\min},\tau_{\max})\) keep the threshold inside \((0,1)\), and the residual cap \(L_h\) limits the history contribution to the acoustic logit.

Automatic initiation requires \(p_t>\tau'_t\) and the state conditions in Section~\ref{study_interface}. The decay constants give interruptions shorter persistence than activations. The history horizon and clipping bound the influence of past feedback on the acoustic decision. Section~\ref{sec:history_selection} compares windows, paths, feature sets, gains, and step sizes.

\subsubsection{Online Updates}

To stabilize learning from sparse, class-imbalanced corrections, we reuse both feedback types through class-balanced replay. We draw on L2-SP's use of pretrained parameters as a regularization reference~\cite{xuhong2018explicit} and FedProx's proximal penalty around the current global model~\cite{li2020federated} to construct a quadratic penalty toward the fixed base head \(\phi_0\), constraining drift during personalization. Here, \(\phi_0\) remains fixed throughout both study stages. \(\phi_{\rm sh}^{*}\) is the averaged starting head locked for new users after Stage~1 (Section~\ref{shared_initialization}), with \(*\) marking its locked status. \(\phi_i\) is the head updated for user \(i\).

To reuse scarce corrections across updates~\cite{rolnick2019experience,ji2021coordinating,liu2024prior}, we store their acoustic representations and labels when scheduling head updates. Separate capacity-limited replay buffers retain past examples for each label. When a buffer is full, reservoir sampling either replaces a randomly selected stored example with the new one or discards the new one. Each update batch uses available examples from both buffers. Losses are averaged within each available class, then across classes, with the current feedback class weighted by \(c_ts_t\) and the other by 1. Here \(c_t={\rm clip}(\sqrt{(n_{\rm opposite}+1)/(n_{\rm same}+1)},c_{\min},c_{\max})\) uses prior correction counts \(n_{\rm same}\) and \(n_{\rm opposite}\) with the same and opposite labels as the current tap. Unit pseudocounts stabilize the ratio for sparse feedback, while fixed bounds \(c_{\min}\) and \(c_{\max}\) limit class reweighting. \(s_t\) is the update scale below. The~objective~is
\[
L_i=L_{\rm replay}+\frac{10^{-4}}{2}\|\phi_i\|_2^2
+\frac{0.01}{2}\|\phi_i-\phi_0\|_2^2.
\]
Here \(L_i\) is the total objective and \(L_{\rm replay}\) the replay loss weighted by class. The two quadratic terms penalize head magnitude and distance from the base head, respectively. With nominal step size \(\eta_{\rm head}\), the implementation scales the gradient step by \(s_t\) and interpolates the resulting parameters toward their pre-update values by the same factor.
\[
\begin{aligned}
\tilde\phi_i&=\phi_i-\eta_{\rm head}s_t\,\bar g_t,\\
\phi_i^+&=\phi_i+s_t(\tilde\phi_i-\phi_i),
\end{aligned}
\]
Here \(\tilde\phi_i\) is the tentative head, \(\phi_i^+\) the updated head, and \(\bar g_t\) the gradient of \(L_i\) with its Euclidean norm clipped. Each scheduled event takes one step. To moderate early head changes, \(s_t\) follows a warm-up schedule based on the number of prior eligible corrections. Calibration bias and history parameters can update during the initial head warm-up. Individual calibration follows
\[
\begin{aligned}
b_i^{{\rm cal},+}={\rm clip}\!\bigl(&b_i^{\rm cal}\\
&+\eta_{\rm cal}\,[c_t(y_t-\hat p_t)-\lambda_{\rm cal}b_i^{\rm cal}],\\
&-B_{\rm cal},B_{\rm cal}\bigr),
\end{aligned}
\]
Here \(b_i^{{\rm cal},+}\) is the updated calibration bias, \(y_t\) the current binary correction label, \(c_t\) the feedback weight, and \(\hat p_t\) the calibrated acoustic probability before updating. The step size \(\eta_{\rm cal}\) controls update magnitude, and shrinkage \(\lambda_{\rm cal}\) pulls the bias toward zero. The bound \(B_{\rm cal}\) limits the calibration contribution to the acoustic logit. Calibration is reset for each participant and excluded from shared-head averaging.

To keep the current tap out of its own history features, the system computes \(x_t\) before adding that tap to history. For acoustic comparisons, the system uses the input paired with that tap's correction label (Section~\ref{framework_overview}). To bound sensitivity to individual history features, we constrain each residual weight to \([-W_h,W_h]\). With history step size \(\eta_h\), an interruption increases \(k_{\rm int}\) and \(b_h\) and updates \(w_h\leftarrow{\rm clip}(w_h-\eta_hx_t,-W_h,W_h)\), making the system more conservative. An activation symmetrically increases \(k_{\rm act}\), decreases \(b_h\), and updates \(w_h\leftarrow{\rm clip}(w_h+\eta_hx_t,-W_h,W_h)\), making it more responsive. Only after these updates is the tap inserted into history for subsequent decisions.

\subsubsection{Illustrative Online Feedback Sequence} \label{sec:feedback_example} Consider an illustrative user in the shared-initialization adaptation condition. The following sequence follows an activation and a later interruption through control, supervision, learning, and subsequent decisions.\par \paragraph{1. Initialization and warm-up.} The user starts from \(\phi_{\rm sh}^{*}\), while the proximal reference remains \(\phi_0\). The encoder and feature normalization stay fixed, individual calibration is reset, and recent correction history starts empty. During the first two eligible corrections, \(s_t=0\): calibration and history parameters can change while the personal head and its replay buffers remain unchanged. The head-update scale subsequently increases according to Section~\ref{implementation_details}. For the two events below, assume that head warm-up has finished and both feedback classes are available for replay.\par \paragraph{2. Activation: link ``respond now'' to the current input.} The user finishes a question, but the system remains in Waiting. An accepted \emph{Answer now} tap immediately starts response generation. The system pairs the audio available at that tap with \(y=1\), extracts its fixed, normalized acoustic representation, and adds it to the positive replay buffer. Using available examples from both classes, it updates the personal head with the class-weighted replay loss, head-magnitude penalty, and proximal penalty toward \(\phi_0\). Gradient clipping and the update scale constrain the step. Thus the question provides positive supervision for starting a response, while replay retains earlier evidence about when to wait.\par \paragraph{3. Calibration and history: prepare the next decision.} For that same activation, the calibrated acoustic probability computed before the update supplies \(\hat p_t\) in the calibration rule. With \(y=1\), its feedback term pushes the calibration bias toward responding, subject to shrinkage and bounds. Before inserting the tap into history, the system forms \(x_t\) from earlier corrections. It increases \(k_{\rm act}\), decreases \(b_h\), and adjusts \(w_h\) in the activation direction. It then appends the tap, so its recency and action become available to subsequent decisions. These history changes favor responding through both the threshold and logit paths.\par \paragraph{4. Interruption: correct the earlier automatic start.} After the manually requested reply has ended, the user begins another utterance. Suppose the system automatically starts responding during a pause before the user has finished. It saves the audio prefix at that automatic start. An accepted \emph{Stop answering} tap in Thinking or Speaking immediately cancels generation or playback and returns the system to Waiting. The negative example pairs \(y=0\) with the saved \emph{start-time} input. This example enters the negative replay buffer and undergoes the same replay and regularized head update. Its calibration feedback favors waiting; the history update increases \(k_{\rm int}\) and \(b_h\) and adjusts \(w_h\) in the interruption direction before appending the tap. The feedback therefore supplies negative supervision for the earlier initiation decision. Canceling a manually requested reply provides immediate stopping control; negative learning examples are formed from accepted interruptions of automatically initiated replies.\par \paragraph{5. Reuse in later interaction.} For each new speech prefix, BPRHT combines the updated head and calibration with the history-dependent logit residual, then compares the resulting probability with the adapted threshold. Automatic response generation starts only when that comparison and the state conditions are satisfied. The activation and interruption thus provide opposing evidence that is reused together to guide subsequent initiation decisions. Recent-event contributions decay and leave the 60\,s history window; learned head, calibration, and history parameters remain available for later decisions and further feedback.
\subsection{Implementation Details}
\label{implementation_details}
\label{dialogue_memory}
This section specifies the basic response-generation and memory services and the evaluation settings for BPRHT (Section~\ref{model_and_learning}). Table~\ref{tab:history_features} defines the history-feature transformations, and Section~\ref{sec:history_selection} compares alternative history configurations.
\paragraph{Speech generation and dialogue memory.}
The response-generation module uses a cascaded ASR--LLM--TTS pipeline. GLM-5.2 generated replies for both the desktop and mobile companion, and the same configured LLM produced the long-term memory summaries. These implementations provide the fixed conversational setting for evaluating BPRHT.\par
For timing inference, the evaluated system continuously processes speech and recent taps. Automatic or manual activation starts the response-generation pipeline. To maintain dialogue continuity, it uses short-term context and asynchronous, nonblocking \emph{Long-term Memory} summaries, drawing on textual episodic memory~\cite{shinn2023reflexion}. To keep dialogue streams separate and context compact, the system stores a length-bounded summary per user and dialogue source. Before reply generation, the summary is read synchronously as system context, with instructions against verbatim repetition. After reply delivery, a background LLM merges the previous summary with the completed turn. Successive requests retain only the latest pending update, and the audit log records summary updates. TTS renders the LLM's recorded reply text to keep the content log consistent with the spoken response. The transition from Waiting to Thinking immediately signals that response preparation has begun.
\paragraph{Decision-head updates.}
Each class-specific replay buffer retains at most 12 examples, and an update batch contains up to 24 available examples. The nominal step size is \(\eta_{\rm head}=0.03\), with gradient-norm clipping at 2. The coefficients of the head-magnitude and base-head penalties in \(L_i\) are \(10^{-4}\) and \(0.01\), respectively. Class reweighting uses \(c_{\min}=0.35\) and \(c_{\max}=3\). In user sessions, the warm-up schedule sets \(s_t=0,0.05,0.2,0.5,1\) for 0--1, 2--3, 4--5, 6--7, or at least 8 prior eligible corrections, respectively.
\paragraph{History settings.}
BPRHT initially assigns a larger gain to interruption feedback, starting with \(k_{\rm int}=0.08\) and \(k_{\rm act}=0.04\). We initialize \(b_h=0\) and \(w_h=\mathbf 0\) so that the bias and residual add no initial offset, and use online step size \(\eta_h=0.05\). The lower modulation bounds are \(\alpha_{\rm ac}=0.55\) and \(\alpha_{\rm ab}=0.60\). Missing-vector fallback values are \(\bar\gamma_{\rm ac}=0.70\) and \(\bar\gamma_{\rm ab}=0.75\), respectively.
Threshold-shift limits are \((\delta_{\rm exp}^{-},\delta_{\rm exp}^{+})=(0.05,0.10)\) before adding \(b_h\) and \((\delta_\tau^{-},\delta_\tau^{+})=(0.06,0.12)\) afterward. Final threshold bounds are \((\tau_{\min},\tau_{\max})=(0.05,0.99)\). The logit-residual cap is \(L_h=0.5\), and residual weights are bounded by \(W_h=0.2\).
\paragraph{Calibration and initiation.}
Settings in the user study were \(T=4\), \(b_g=0.8\), \(b_i^{\rm cal}=0\) initially, and \(\tau_0=0.70\). Individual calibration uses step size \(\eta_{\rm cal}=0.25\), shrinkage \(\lambda_{\rm cal}=0.05\), and bias bound \(B_{\rm cal}=1.25\).
\paragraph{Base initialization.}
To enable BPRHT to make response decisions before receiving online corrections, we fit its base head and feature normalization on 300 synthetic development examples from 12 simulated profiles, comprising 132 positive and 168 negative examples, using standardized logistic regression with inverse regularization strength 0.1. This head supplies \(\phi_0\), the fixed reference used by the quadratic penalty to limit parameter drift during personalization.
\FloatBarrier
\section{Technical Evaluation}
\setcounter{dbltopnumber}{1}
\label{Experiment}
\label{val_exp_section}

\begin{table*}[t]
\centering
\caption{Acoustic model comparison on 1,000 recordings using five seeds. Left: mean $\pm$ sample SD, with initial AUPRC on samples 1--200 and final metrics on 801--1,000. Right: F1 and Brier score in 200-sample windows with 100-sample overlap. Points are five-seed means joined by lines, with row symbols identifying models. Dashed callouts identify the selected DistilHuBERT backbone. P95 is waveform-to-probability latency (ms). $\dagger$ marks simplified mechanism-based controls. Bold marks the best final means. ECE: expected calibration error. DTCN: dilated temporal convolutional network.}
\label{tab:offline_model_selection}
\begin{minipage}[t]{0.55\linewidth}\vspace{0pt}\centering\fontsize{7.0}{8.4}\selectfont
\setlength{\tabcolsep}{1.1pt}\renewcommand{\arraystretch}{1.04}\tagpdfsetup{table/header-rows={1,2}}
\begin{tabular*}{\linewidth}{@{\extracolsep{\fill}}lrrrrrr@{}}
\toprule
Model & \shortstack{1--200\\AUPRC} & \multicolumn{4}{c}{Samples 801--1,000} & \shortstack{P95\\(ms)} \\ \cmidrule(lr){3-6} & & F1 & AUPRC & Brier\(\downarrow\) & ECE\(\downarrow\) & \\
\midrule
\includegraphics[width=8pt]{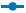}\,DistilHuBERT & \shortstack[r]{0.6913\\\(\pm 0.0124\)} & \shortstack[r]{\textbf{0.8942}\\\(\pm 0.0219\)} & \shortstack[r]{\textbf{0.9627}\\\(\pm 0.0102\)} & \shortstack[r]{0.0500\\\(\pm 0.0051\)} & \shortstack[r]{0.0755\\\(\pm 0.0016\)} & 22.14 \\
\modelrowseparator
\includegraphics[width=8pt]{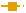}\,\shortstack[l]{Acoustic +\\DistilHuBERT} & \shortstack[r]{0.7425\\\(\pm 0.0133\)} & \shortstack[r]{0.8788\\\(\pm 0.0126\)} & \shortstack[r]{0.9532\\\(\pm 0.0111\)} & \shortstack[r]{\textbf{0.0478}\\\(\pm 0.0043\)} & \shortstack[r]{0.0565\\\(\pm 0.0080\)} & 29.25 \\
\modelrowseparator
\includegraphics[width=8pt]{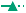}\,\shortstack[l]{Whisper tiny\\encoder} & \shortstack[r]{0.8240\\\(\pm 0.0129\)} & \shortstack[r]{0.8499\\\(\pm 0.0079\)} & \shortstack[r]{0.9573\\\(\pm 0.0053\)} & \shortstack[r]{0.0600\\\(\pm 0.0027\)} & \shortstack[r]{0.0981\\\(\pm 0.0081\)} & 11.84 \\
\modelrowseparator
\includegraphics[width=8pt]{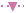}\,HuBERT base & \shortstack[r]{0.6719\\\(\pm 0.0171\)} & \shortstack[r]{0.8744\\\(\pm 0.0084\)} & \shortstack[r]{0.9526\\\(\pm 0.0084\)} & \shortstack[r]{0.0535\\\(\pm 0.0034\)} & \shortstack[r]{0.0709\\\(\pm 0.0022\)} & 37.89 \\
\modelrowseparator
\includegraphics[width=8pt]{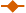}\,BEATs & \shortstack[r]{0.7673\\\(\pm 0.0123\)} & \shortstack[r]{0.8769\\\(\pm 0.0123\)} & \shortstack[r]{0.9514\\\(\pm 0.0085\)} & \shortstack[r]{0.0514\\\(\pm 0.0035\)} & \shortstack[r]{0.0742\\\(\pm 0.0063\)} & 49.98 \\
\modelrowseparator
\includegraphics[width=8pt]{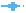}\,WavLM layer 6 & \shortstack[r]{0.7349\\\(\pm 0.0235\)} & \shortstack[r]{0.8176\\\(\pm 0.0182\)} & \shortstack[r]{0.9402\\\(\pm 0.0096\)} & \shortstack[r]{0.0669\\\(\pm 0.0040\)} & \shortstack[r]{0.0908\\\(\pm 0.0057\)} & 53.15 \\
\modelrowseparator
\includegraphics[width=8pt]{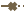}\,\shortstack[l]{Acoustic\\statistics} & \shortstack[r]{0.7022\\\(\pm 0.0065\)} & \shortstack[r]{0.8626\\\(\pm 0.0102\)} & \shortstack[r]{0.9344\\\(\pm 0.0076\)} & \shortstack[r]{0.0571\\\(\pm 0.0012\)} & \shortstack[r]{0.0593\\\(\pm 0.0037\)} & 9.69 \\
\modelrowseparator
\includegraphics[width=8pt]{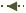}\,\shortstack[l]{Wav2Vec2\\emotion} & \shortstack[r]{0.5826\\\(\pm 0.0089\)} & \shortstack[r]{0.8613\\\(\pm 0.0158\)} & \shortstack[r]{0.9294\\\(\pm 0.0132\)} & \shortstack[r]{0.0622\\\(\pm 0.0047\)} & \shortstack[r]{0.0776\\\(\pm 0.0027\)} & 38.92 \\
\modelrowseparator
\includegraphics[width=8pt]{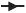}\,DTCN~\cite{lea2017temporal} & \shortstack[r]{0.5717\\\(\pm 0.0361\)} & \shortstack[r]{0.8805\\\(\pm 0.0188\)} & \shortstack[r]{0.9086\\\(\pm 0.0444\)} & \shortstack[r]{0.0601\\\(\pm 0.0125\)} & \shortstack[r]{\textbf{0.0553}\\\(\pm 0.0232\)} & 6.00 \\
\modelrowseparator
\includegraphics[width=8pt]{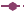}\,Jasper~\cite{li2019jasper} & \shortstack[r]{0.3255\\\(\pm 0.0306\)} & \shortstack[r]{0.7236\\\(\pm 0.0470\)} & \shortstack[r]{0.7180\\\(\pm 0.1019\)} & \shortstack[r]{0.1546\\\(\pm 0.0397\)} & \shortstack[r]{0.1510\\\(\pm 0.0440\)} & 11.56 \\
\modelrowseparator
\includegraphics[width=8pt]{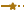}\,Conformer~\cite{gulati2020conformer} & \shortstack[r]{0.4426\\\(\pm 0.0460\)} & \shortstack[r]{0.8693\\\(\pm 0.0247\)} & \shortstack[r]{0.8477\\\(\pm 0.0392\)} & \shortstack[r]{0.0667\\\(\pm 0.0136\)} & \shortstack[r]{0.0564\\\(\pm 0.0126\)} & 22.44 \\
\modelrowseparator
\includegraphics[width=8pt]{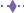}\,\shortstack[l]{Skantze-style\\LSTM\(^{\dagger}\)~\cite{skantze2017towards}} & \shortstack[r]{0.3351\\\(\pm 0.0221\)} & \shortstack[r]{0.5007\\\(\pm 0.0392\)} & \shortstack[r]{0.4511\\\(\pm 0.0395\)} & \shortstack[r]{0.2025\\\(\pm 0.0084\)} & \shortstack[r]{0.1785\\\(\pm 0.0271\)} & 2.57 \\
\modelrowseparator
\includegraphics[width=8pt]{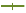}\,VAP-style\(^{\dagger}\)~\cite{ekstedt2022voice} & \shortstack[r]{0.3753\\\(\pm 0.0181\)} & \shortstack[r]{0.5108\\\(\pm 0.0125\)} & \shortstack[r]{0.5129\\\(\pm 0.0178\)} & \shortstack[r]{0.2119\\\(\pm 0.0017\)} & \shortstack[r]{0.2112\\\(\pm 0.0027\)} & 6.71 \\
\modelrowseparator
\includegraphics[width=8pt]{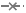}\,\shortstack[l]{ICARUS-style\\GMM\(^{\dagger}\)~\cite{li2022can}} & \shortstack[r]{0.3240\\\(\pm 0.0118\)} & \shortstack[r]{0.4540\\\(\pm 0.0781\)} & \shortstack[r]{0.3951\\\(\pm 0.0863\)} & \shortstack[r]{0.3014\\\(\pm 0.0323\)} & \shortstack[r]{0.2912\\\(\pm 0.0357\)} & 7.42 \\
\bottomrule
\end{tabular*}\end{minipage}\hfill\begin{minipage}[t]{0.435\linewidth}\vspace{0pt}\centering\includegraphics[width=\linewidth,alt={F1 and Brier score for fourteen models in 200-sample windows advanced by 100 samples. Points are five-seed means. Dashed callouts identify DistilHuBERT, with other curves shown more lightly. Symbols correspond to table rows.}]{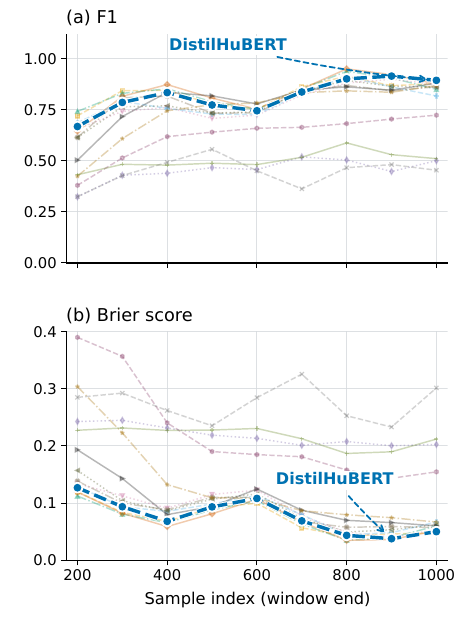}\end{minipage}
\end{table*}

To establish a technical basis for the user study, we assess the technical performance of the response-initiation loop implemented within the Tap-to-Adapt framework and select its algorithmic configuration through two complementary evaluations. Offline evaluation on 1,000 manually labeled recordings assesses response classification, probability calibration, and inference latency. Simulation evaluation uses fixed LLM+TTS speech streams to exercise the control-and-learning loop and compare acoustic representations, online update rules, and recent correction history under shared speech inputs and response opportunities.

\begin{table*}[t]
\centering
\caption{Update-rule and acoustic-representation comparisons on fixed simulated speech streams. Correction and interruption rates use all opportunities, while activation uses response-required opportunities. Final-segment metrics cover the last 18 minutes per profile.}
\label{tab:trainer_selection}
\small
\setlength{\tabcolsep}{3pt}
\begin{tabular*}{\linewidth}{@{\extracolsep{\fill}}lrrrr@{}}
\toprule
Candidate & \shortstack{Overall\\correction\(\downarrow\)} & \shortstack{Final-segment\\correction\(\downarrow\)} & \shortstack{Manual\\activation} & \shortstack{Manual\\interruption} \\
\midrule
\multicolumn{5}{@{}l}{\textit{Training configurations (DistilHuBERT reference and feature controls)}} \\
DistilHuBERT + balanced proximal replay & \textbf{0.0188} & \textbf{0.0139} & \textbf{0.0271} & 0.0052 \\
DistilHuBERT + gated replay & 0.0250 & 0.0174 & 0.0396 & 0.0052 \\
DistilHuBERT + stabilized replay & 0.0302 & 0.0208 & 0.0500 & 0.0052 \\
Acoustic statistics + stabilized replay & 0.0604 & 0.0590 & 0.1104 & 0.0052 \\
Acoustic + DistilHuBERT + stabilized replay & 0.0896 & 0.0764 & 0.1688 & 0.0052 \\
DistilHuBERT + frozen user-specific head & 0.1104 & 0.1042 & 0.2104 & 0.0052 \\
\addlinespace
\multicolumn{5}{@{}l}{\textit{Backbone ablation with fixed balanced proximal replay; the DistilHuBERT row above is the reference}} \\
Acoustic + DistilHuBERT + balanced proximal replay & \textbf{0.0188} & \textbf{0.0139} & \textbf{0.0271} & 0.0052 \\
DTCN + balanced proximal replay~\cite{lea2017temporal} & 0.0677 & 0.0208 & 0.1250 & 0.0052 \\
\bottomrule
\end{tabular*}
\end{table*}

\begin{table*}[t]
\centering
\caption{Recent-history configuration comparisons. Cells show mean $\pm$ sample SD across five shared seeds. $\Delta$ is the paired difference from B. Correction and interruption use all opportunities, while activation uses response-required opportunities. Bold marks the lowest correction and interruption means and the most negative $\Delta$. B denotes no recent history, A1 the reference configuration, and A10 the selected BPRHT configuration.}
\label{tab:history_ablation}
\small
\setlength{\tabcolsep}{3pt}
\begin{tabular*}{\textwidth}{@{\extracolsep{\fill}}c p{0.43\textwidth} rrrr@{}}
\toprule
ID & Configuration & \shortstack{Overall\\corr.\(\downarrow\)} & \shortstack{Manual\\activation} & \shortstack{Manual\\interruption\(\downarrow\)} & \shortstack{\(\Delta\)corr.\\vs B} \\
\midrule
B & BPRHT without recent history; history-dependent threshold shift and logit residual set to zero & \shortstack[r]{0.1065\\{\footnotesize\(\pm 0.0140\)}} & \shortstack[r]{0.0271\\{\footnotesize\(\pm 0.0000\)}} & \shortstack[r]{0.0929\\{\footnotesize\(\pm 0.0140\)}} & --- \\
\historyrowseparator
A1 & Full history features; 60\,s trailing window; threshold shift and bounded logit residual; \(\eta_h=.025\) & \shortstack[r]{0.0838\\{\footnotesize\(\pm 0.0070\)}} & \shortstack[r]{0.0304\\{\footnotesize\(\pm 0.0019\)}} & \shortstack[r]{0.0685\\{\footnotesize\(\pm 0.0075\)}} & \shortstack[r]{-0.0227\\{\footnotesize\(\pm 0.0076\)}} \\
\historyrowseparator
A2 & Reference full features and 60\,s window; threshold-shift path only & \shortstack[r]{0.0877\\{\footnotesize\(\pm 0.0080\)}} & \shortstack[r]{0.0300\\{\footnotesize\(\pm 0.0011\)}} & \shortstack[r]{0.0727\\{\footnotesize\(\pm 0.0084\)}} & \shortstack[r]{-0.0188\\{\footnotesize\(\pm 0.0071\)}} \\
\historyrowseparator
A3 & Reference full features and 60\,s window; bounded logit-residual path only & \shortstack[r]{0.1006\\{\footnotesize\(\pm 0.0135\)}} & \shortstack[r]{0.0279\\{\footnotesize\(\pm 0.0011\)}} & \shortstack[r]{0.0867\\{\footnotesize\(\pm 0.0139\)}} & \shortstack[r]{-0.0058\\{\footnotesize\(\pm 0.0048\)}} \\
\historyrowseparator
A4 & Full features and 60\,s window; logit-clipping fusion & \shortstack[r]{0.1002\\{\footnotesize\(\pm 0.0130\)}} & \shortstack[r]{0.0283\\{\footnotesize\(\pm 0.0019\)}} & \shortstack[r]{0.0860\\{\footnotesize\(\pm 0.0134\)}} & \shortstack[r]{-0.0063\\{\footnotesize\(\pm 0.0065\)}} \\
\historyrowseparator
A5 & 30\,s trailing history window; otherwise the reference & \shortstack[r]{0.0842\\{\footnotesize\(\pm 0.0069\)}} & \shortstack[r]{0.0308\\{\footnotesize\(\pm 0.0017\)}} & \shortstack[r]{0.0688\\{\footnotesize\(\pm 0.0075\)}} & \shortstack[r]{-0.0223\\{\footnotesize\(\pm 0.0089\)}} \\
\historyrowseparator
A6 & 90\,s trailing history window; otherwise the reference & \shortstack[r]{\textbf{0.0831}\\{\footnotesize\(\pm 0.0067\)}} & \shortstack[r]{0.0304\\{\footnotesize\(\pm 0.0019\)}} & \shortstack[r]{\textbf{0.0679}\\{\footnotesize\(\pm 0.0072\)}} & \shortstack[r]{\textbf{-0.0233}\\{\footnotesize\(\pm 0.0081\)}} \\
\historyrowseparator
A7 & 60\,s, both paths; temporal features only, including time since taps & \shortstack[r]{0.0842\\{\footnotesize\(\pm 0.0075\)}} & \shortstack[r]{0.0304\\{\footnotesize\(\pm 0.0019\)}} & \shortstack[r]{0.0690\\{\footnotesize\(\pm 0.0079\)}} & \shortstack[r]{-0.0223\\{\footnotesize\(\pm 0.0068\)}} \\
\historyrowseparator
A8 & 60\,s, both paths; minimal history features & \shortstack[r]{0.0879\\{\footnotesize\(\pm 0.0081\)}} & \shortstack[r]{0.0300\\{\footnotesize\(\pm 0.0011\)}} & \shortstack[r]{0.0729\\{\footnotesize\(\pm 0.0085\)}} & \shortstack[r]{-0.0185\\{\footnotesize\(\pm 0.0066\)}} \\
\historyrowseparator
A9 & Reference plus two quarter-window decay slots; all four reported metrics match the reference & \shortstack[r]{0.0838\\{\footnotesize\(\pm 0.0070\)}} & \shortstack[r]{0.0304\\{\footnotesize\(\pm 0.0019\)}} & \shortstack[r]{0.0685\\{\footnotesize\(\pm 0.0075\)}} & \shortstack[r]{-0.0227\\{\footnotesize\(\pm 0.0076\)}} \\
\historyrowseparator
A10 & Full features, 60\,s, both paths; online step size \(\eta_h=.05\) (BPRHT) & \shortstack[r]{0.0844\\{\footnotesize\(\pm 0.0088\)}} & \shortstack[r]{0.0329\\{\footnotesize\(\pm 0.0017\)}} & \shortstack[r]{\textbf{0.0679}\\{\footnotesize\(\pm 0.0092\)}} & \shortstack[r]{-0.0221\\{\footnotesize\(\pm 0.0060\)}} \\
\historyrowseparator
A11 & Online step size \(\eta_h=.01\); otherwise the reference & \shortstack[r]{0.0908\\{\footnotesize\(\pm 0.0113\)}} & \shortstack[r]{0.0292\\{\footnotesize\(\pm 0.0015\)}} & \shortstack[r]{0.0763\\{\footnotesize\(\pm 0.0113\)}} & \shortstack[r]{-0.0156\\{\footnotesize\(\pm 0.0068\)}} \\
\historyrowseparator
A12 & Initial gains \(k_{\rm int0}=.12,\ k_{\rm act0}=.03\); otherwise the reference & \shortstack[r]{0.0833\\{\footnotesize\(\pm 0.0066\)}} & \shortstack[r]{0.0304\\{\footnotesize\(\pm 0.0019\)}} & \shortstack[r]{0.0681\\{\footnotesize\(\pm 0.0070\)}} & \shortstack[r]{-0.0231\\{\footnotesize\(\pm 0.0080\)}} \\
\historyrowseparator
A13 & Minimal history features; threshold-shift path only & \shortstack[r]{0.0883\\{\footnotesize\(\pm 0.0078\)}} & \shortstack[r]{0.0300\\{\footnotesize\(\pm 0.0011\)}} & \shortstack[r]{0.0733\\{\footnotesize\(\pm 0.0082\)}} & \shortstack[r]{-0.0181\\{\footnotesize\(\pm 0.0067\)}} \\
\historyrowseparator
A14 & Temporal features only; bounded logit-residual path only & \shortstack[r]{0.1025\\{\footnotesize\(\pm 0.0139\)}} & \shortstack[r]{0.0279\\{\footnotesize\(\pm 0.0011\)}} & \shortstack[r]{0.0885\\{\footnotesize\(\pm 0.0143\)}} & \shortstack[r]{-0.0040\\{\footnotesize\(\pm 0.0038\)}} \\
\bottomrule
\end{tabular*}
\end{table*}
\subsection{Evaluation on Recorded Speech}

For selection of the acoustic backbone, we recorded scripted interactions with the prototype in annotation mode. Scenarios varied the topic (ten DailyDialog categories~\cite{li2017dailydialog}), speaking distance (30, 60, or 90\,cm), prior dialogue context (present or absent), addressee (agent or nearby person), and turn state. Within a sequence, the topic, distance, and addressee were each retained from the preceding scenario with probability 0.5. A completed turn with a reply expected had a sampling probability of 0.5. The remaining probability was divided equally among ending a turn without expecting a reply, pausing during speech, speaking without a pause, and silence. The intended \emph{begin responding} condition was a completed turn addressed to the agent for which a reply was expected. The first author led collection and assigned most labels, with an external collaborator assisting collection. Each annotation retained the preceding 15\,s of mono audio. We used the last 1,000 recordings in collection order, with 253 \emph{begin responding} and 747 \emph{continue waiting} labels.

Candidates predicted before updating in collection order, using five fixed random seeds shared across candidates and classification threshold \(\tau=0.5\) as a common binary-classification operating point for this recorded-speech comparison. We summarize learning in 200-sample windows advanced by 100 samples, and compare initial samples 1--200 with final samples 801--1,000. Table~\ref{tab:offline_model_selection} reports all candidates and their windowed F1 and Brier trajectories. Means and sample standard deviations describe training variability across seeds on the same stream. F1 is the harmonic mean of precision and recall for \emph{begin responding}. AUPRC is computed as average precision for the same positive class. The Brier score is the mean squared error between predicted response probabilities and binary labels~\cite{glenn1950verification}. ECE is the sample-weighted mean absolute gap between mean predicted probability and positive-label frequency across ten equal-frequency bins.

Pretrained representations remain frozen and are individually \(\ell_2\)-normalized to keep input scales comparable. Fusion first normalizes acoustic statistics and DistilHuBERT features separately. Concatenating two unit-norm blocks gives norm \(\sqrt2\), so division by this value restores the unit norm used by the single-representation inputs. The frozen-feature candidates share an online logistic learner to keep adaptation settings consistent across representations. Its settings were fixed in an earlier development pass on samples 101--600: learning rate 0.03 controls the update step, \(\ell_2\) coefficient \(10^{-4}\) shrinks head weights, and at most 12 replay samples per class reuse feedback with bounded memory. Each feedback event triggers one update.\par
Jasper~\cite{li2019jasper} and Conformer~\cite{gulati2020conformer} use the same 1,000 recordings, resampling 15-second clips to 4\,kHz, right-padding to 30 seconds, and extracting 40-dimensional Mel inputs. Both learn the binary labels using Adam with a learning rate of \(5\times10^{-5}\), class-balanced replay from the latest 200 samples, six updates per event, and gradient-norm clipping at 1. Jasper has ten convolutional blocks. Conformer has 12 encoder blocks, 256-dimensional hidden representations, four attention heads, and fourfold downsampling.\par
Table~\ref{tab:offline_model_selection} also includes three mechanism-based controls informed by turn-taking research. Each follows the same five-seed predict-then-update protocol on the 1,000-sample stream. Skantze-style LSTM~\cite{skantze2017towards} uses 40-dimensional Mel sequences, one 64-unit LSTM layer, temporal mean pooling, and a binary linear head, with the same preprocessing, Adam configuration, and six balanced-replay updates as Jasper and Conformer. VAP-style~\cite{ekstedt2022voice} uses trailing silence, voice-activity proportion, and energy and zero-crossing statistics over the latest 1, 0.5, and 0.25 seconds. Inputs receive per-sample \(\ell_2\) normalization and the same online logistic learner as the frozen-feature models. ICARUS-style GMM~\cite{li2022can} adds whole-clip energy and zero-crossing statistics and fits a two-component diagonal-covariance GMM per class. Starting after sample 40, normalization, class-conditional densities, and smoothed priors are re-estimated every 20 samples from at most the latest 200 labeled examples, once each class contains at least eight samples. Before the first fit, prediction uses a Beta(1,1)-smoothed prior. The two statistical controls compute acoustic statistics on the CPU. All three controls target this task's respond and wait labels. Their results characterize these simplified Mel-LSTM, voice-activity-statistics, and class-conditional GMM mechanisms. The original systems are described in Section~\ref{related_work_speech_turn_init}.

We measure single-example latency on an RTX 4070 Laptop GPU with models trained using a fixed random seed and frozen after 1,000 updates. From 30 recordings evenly spaced in collection order, the first five warm up the pipeline. All 30 are then timed individually in that order with batch size 1. Timing spans in-memory 15-second mono 16\,kHz waveform input to response probability returned to the CPU, including preprocessing, transfers, encoding, pooling, normalization, and the head, with GPU synchronization at both boundaries. It excludes file I/O, model loading, updates, history conditioning, and dialogue generation. CPU and numerical-library threads are fixed at two. Neural models use float16 mixed precision. Acoustic statistics and statistical controls run on the CPU. Whisper uses only its encoder. WavLM selects layer 6. P95, the linearly interpolated empirical 95th percentile of 30 measurements, meets the deployment criterion of \(\leq100\) ms for all candidates.

DistilHuBERT achieved final F1 \(0.8942\pm0.0219\) and AUPRC \(0.9627\pm0.0102\), the highest means among the candidates. Its P95 latency from waveform input to response probability was 22.14\,ms on an RTX 4070 Laptop GPU. All candidates met the 100\,ms deployment criterion. These results support DistilHuBERT's accuracy within the timing budget.

\subsection{Simulation Evaluation}
\label{sec:history_selection}

We use simulated speech interaction scenarios to evaluate and select acoustic representations, online update rules, and settings for using recent history in BPRHT (Section~\ref{model_and_learning}). Fixed dialogue streams provide common speech inputs and response expectations for comparing candidate configurations.

\paragraph{Simulated Interaction Scenarios.}
We generated dialogue scripts with GLM-5.2 from prespecified topics, speaker roles, and response intentions. We synthesize fixed mono 16\,kHz PCM using sherpa-onnx with the AISHELL3 VITS model. A \emph{simulated speaker profile} pairs a fixed synthetic user voice with a pre-generated dialogue script. The eight profiles vary in voice identity and topic sequence while sharing the same interaction structure. Each six-minute cycle contains six utterances addressed to the experimental agent with a reply expected, two with no reply expected, two addressed to the companion, and two companion replies. The voices of the experimental agent, companion agent, and simulated users do not overlap. Each candidate maintains a separate decision head per profile, initialized from a common head fitted on separate LLM+TTS development profiles. Predictions precede feedback-driven updates. Within each comparison, candidates share dialogue text, audio, response expectations, state machine, and update opportunities.\par
The baseline threshold \(\tau_0\) is the response-probability cutoff before history-dependent threshold shifts (Section~\ref{model_and_learning}). Its simulation value, \(\tau_0=0.94\), was selected on six held-out synthetic profiles (150 examples) using balanced accuracy under response-rate constraints and is fixed across candidates. The user study uses \(\tau_0=0.70\) (Section~\ref{implementation_details}).\par
A simulation opportunity is a scripted speech event with a specified addressee and reply expectation. The model evaluates successive speech prefixes at 200\,ms intervals within each opportunity. The streaming cache contains 40,157 prefixes. Evaluation uses the first ten six-minute cycles of each profile, each cycle comprising four minutes of conversation with the experimental agent followed by two minutes with the companion agent. This gives 60 minutes per profile and 960 opportunities across the eight profiles for each condition. The simulator generates an interruption for a premature or unneeded response and a manual activation when an expected response misses the deadline of 1 to 3\,s specified for that scenario. In Tables~\ref{tab:trainer_selection} and~\ref{tab:history_ablation}, overall correction and interruption rates use all simulated opportunities, while manual activation rates use those requiring a response.

\paragraph{Update Rules and Acoustic Representations.}
To assess how online update rules affect correction burden, we compare the DistilHuBERT configurations using balanced proximal replay, gated replay, stabilized replay, or a frozen head in Table~\ref{tab:trainer_selection}. The table's upper part also includes controls that use acoustic statistics or fused representations under stabilized replay. With balanced proximal replay fixed, the table's lower part tests whether fused or DTCN representations improve on DistilHuBERT. These comparisons cover six configurations of update rules and acoustic features and two further backbone configurations. The remaining backbones are evaluated on the recorded-speech stream in Table~\ref{tab:offline_model_selection}.

To moderate updates supported by only one feedback class, gated replay sets \(s_t=0.75\) when the replay buffer contains only that class. Applying this scale to both the gradient step and parameter interpolation yields the effective step multiplier \(s_t^2=0.75^2\). Stabilized replay limits parameter drift with a learning rate of 0.01, 32 samples per class, and a proximal coefficient of 0.05. The balanced proximal replay reference uses 0.03, 12, and 0.01, respectively.

DistilHuBERT with balanced proximal replay yielded an overall correction rate of 0.0188, compared with 0.0250 for gated replay, 0.0302 for stabilized replay, and 0.1104 for a frozen head. Balanced proximal replay therefore had the lowest overall correction rate among these DistilHuBERT update configurations. With this update rule fixed, adding acoustic statistics yielded the same overall and final-segment correction rates, manual activation rate, and manual interruption rate as DistilHuBERT alone. Brier scores were 0.0107 for the fused representation and 0.0110 for DistilHuBERT alone. The matching decision rates favor the lower-dimensional DistilHuBERT representation. BPRHT uses this frozen representation and balanced proximal replay to train a separate decision head for each user. Updates to decision heads in the user study follow the warm-up schedule in Section~\ref{implementation_details}.

\paragraph{Recent Correction History.}
To assess whether recent corrections help under recurring timing mismatches, we hold DistilHuBERT and balanced proximal replay fixed and compare configurations for using recent history. Logit offsets applied across blocks make the simulated model more or less likely to initiate a response. The simulator also permits initiation at selected simulated pauses within unfinished utterances, creating opportunities for premature responses. Candidates share these perturbations within each run, using the same eight profiles, voices, and synthesis and replay procedure.

Table~\ref{tab:history_ablation} uses B to identify the baseline without recent history and A1--A14 to identify 14 configurations that incorporate recent correction history. B sets the history-dependent threshold shift and logit residual to zero. A1 provides the reference configuration: the full 20-dimensional history features, a 60\,s trailing window, both threshold-shift and bounded logit-residual paths, and online step size \(\eta_h=.025\). Its reference settings include 25\,s and 40\,s constants controlling feedback decay, a 3000\,ms delay scale, a scale for feature differences of 8, default values for missing data, initial gains, and clipping bounds, as specified in Table~\ref{tab:history_features} and Section~\ref{implementation_details}. Each variant changes the factors specified in the table. We compare mean total correction, interruption, and activation rates, in that order.

\begin{figure*}[t]
 \centering
  \includegraphics[width=\linewidth,alt={Three plots compare BPRHT with the no-history baseline in six-minute windows over 60 minutes. Points are five-seed means and curves summarize moving averages. The panels show correction, automatic activation, and manual activation rates. Mean total correction is 0.0844 for BPRHT and 0.1065 without history.}]{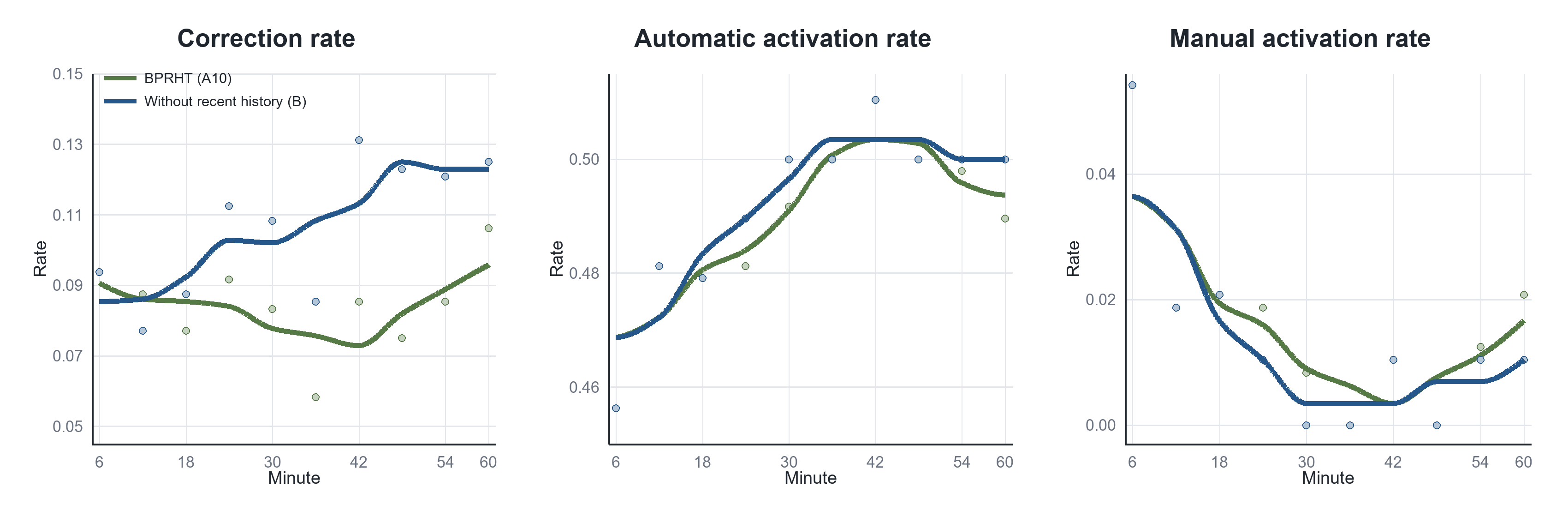}
 \caption{Simulation trajectories for BPRHT (configuration A10) and its baseline without recent history (B), over 60 minutes per profile. Points show five-seed means in six-minute windows. All rates use all opportunities. Curves use shape-preserving cubic interpolation of three-point moving averages, with two points at the endpoints.}
 \Description{Three panels plot Correction rate, Automatic activation rate, and Manual activation rate against elapsed minutes. BPRHT has lower correction rates than the no-history baseline across most windows. Both show increasing automatic activation before leveling off. Manual activation initially declines and then rises slightly. Points show unsmoothed window means. Thick lines summarize local trends.}
 \label{fig:val_exp}
\end{figure*}

Feature comparisons assess how much correction-history information is needed for timing decisions. Temporal features retain feedback timing and action information, with acoustic similarity and feature change marked as missing. The minimal configuration retains decayed activation and interruption strengths, the latest action, whether recent feedback exists, and indicators for missing activation or interruption events. Other variants change the history window, decision paths, initial feedback gains, or online step size.

The initial exploratory run with a fixed random seed selected BPRHT by total correction, breaking ties by manual interruption rate. BPRHT uses the full 20-dimensional feature set, a 60\,s window, both threshold-shift and bounded logit-residual paths, and \(\eta_h=.05\). To examine sensitivity to the simulated perturbations, we evaluated the configurations with five fixed random seeds, varying logit offsets across blocks and response opportunities at pauses within utterances. Dialogue text, cached audio, acoustic features, and each configuration's initial parameters remained fixed across seeds. Candidates shared perturbations and opportunities within each seed and predicted before updating. Means and sample SDs across seeds include the selection seed. Paired differences compare each configuration with the no-history baseline within the same seed.

Across five seeds, BPRHT has total correction \(0.0844\pm0.0088\), compared with \(0.1065\pm0.0140\) for the configuration without recent history, a paired difference of \(-0.0221\pm0.0060\). Across all five runs, BPRHT yielded lower overall correction and interruption rates than the configuration without recent history. Mean interruption rates were 0.0679 and 0.0929, respectively. The 90-second history-window configuration has the lowest mean total correction of 0.0831. BPRHT is close to the temporal-feature configuration at 0.0842. BPRHT and the 90-second configuration tie for the lowest mean interruption rate of 0.0679. Thirteen configurations reduced mean total correction by at least 0.005 relative to the no-history baseline and also had lower mean interruption rates, across the tested history windows, feature sets, and step sizes.

Adding two quarter-window decay channels extends the reference configuration. Its four reported decision metrics are identical to the reference across the five-seed aggregate, indicating no observable decision benefit from these additional channels with the fixed stimuli and threshold. Figure~\ref{fig:val_exp} shows the mean trajectories for BPRHT and its configuration without recent history in six-minute windows. All aggregate metrics use raw observations. Continuous curves summarize trends across these windows.

\section{User Study}
\label{usa_exp}

Following the technical evaluation and algorithm selection in Section~\ref{val_exp_section}, we examine the effects of the Tap-to-Adapt framework in live speech interaction with users. This section addresses two questions. RQ1 asks how shared initialization affects new users' cold-start correction burden and how this effect changes over interaction. RQ2 asks whether users and the system exhibit mutual adaptation around response initiation timing.\par
To address these questions, we conducted a two-stage study. Stage~1 established shared initialization through sequential participation and a prespecified plateau rule, stopping after 20 participants. Stage~2 evaluated it with 48 new users assigned to \emph{cold-start online adaptation}, \emph{shared-initialization online adaptation}, or \emph{frozen after shared initialization}. Comparing the two adapting conditions addresses RQ1. The comparison between shared-initialization adaptation and freezing, feedback-driven update logs, and reported behavioral adjustments inform RQ2. Both stages used five six-minute blocks of desktop and mobile conversation, with corrective taps and a final questionnaire. Participants were the statistical units. Taps and audio frames were repeated observations within participants.
\subsection{Common Protocol and Measures}

\subsubsection{Participants and Procedure}

We recruited participants through instant messaging tools, acquaintance referrals, and online announcements in campus media. Participants could switch between Chinese and English configurations before the session, and the selected language remained fixed throughout the session. Sessions took place primarily on a university campus in China. Sixty-six participants chose Chinese and two chose English. Participants provided informed consent. Declining optional raw-audio retention did not affect participation. Both stages followed the five-block protocol in Section~\ref{preliminary_duration}, with 30 active minutes over 34 minutes including breaks. Shared prompts based on DailyDialog~\cite{li2017dailydialog} supported sustained conversation. Participants could freely choose, develop, and switch topics. Before formal sessions, participants passed checks of device quality and six scenarios to assess understanding of the controls. Restarts after disconnection, backgrounding, or TTS failure cleared segment memory and short-term context and were audited to ensure consistent initial states.

Inclusion required the server to confirm completion of the desktop timed session, mobile companion session, and final questionnaire under the same identifier for the study session. Sessions that participants reported as unsuccessful because of network or other technical failures were excluded from analysis.

\subsubsection{Active Interaction Duration and Blocks}
\label{preliminary_duration}

The protocol specifies 30 minutes of active interaction to observe opening and subsequent adaptation within a fixed budget while balancing task completion, fatigue, and between-condition comparison. Each session has five six-minute active blocks, each comprising four minutes of primary computer conversation followed by two minutes with the push-to-talk mobile companion. A one-minute break follows each of the first four blocks. Timed interaction ends after block 5. The complete schedule is therefore 34 minutes, comprising 20 minutes of computer conversation, ten minutes of mobile conversation, and four minutes of breaks. Computer-side acoustic capture, timing inference, and correction logging span both conversation phases, covering all 30 active minutes. During breaks, audio capture, inference, tap acceptance, and weight updates pause. Analyses use active duration. Device checks, instructions, and the final questionnaire occur outside the 34-minute schedule. Both user-study stages use this configuration.

\subsubsection{Study System, Interfaces, and Feedback Operations}
\label{study_interface}

Participants used a computer and a phone running coordinated desktop and mobile companion clients, supported by backend services. The desktop client ran an implementation of the Tap-to-Adapt framework with BPRHT and the ASR--LLM--TTS generation backend, exchanged continuous 16\,kHz PCM audio, model states, and tap events over a low-latency bidirectional channel, and handled the main conversation and speech recognition. To introduce addressee changes and speech not directed to the desktop, the mobile client provided a push-to-talk (PTT) companion agent. It also received phase instructions and played a fixed environmental sound track. Session, LLM, TTS, questionnaire, and audit services ran in separate processes from the real-time audio service. Production and test instances used separate database, protocol, model, and event namespaces. In the study deployment, a dedicated thread pool processed a 40\,s circular buffer for low-latency timing inference at approximately 180--400\,ms intervals. Each LLM request retained six in-process messages as short-term context. The two clients separately used the independent asynchronous segment summaries described in Section~\ref{dialogue_memory} to maintain multi-turn coherence.

Across cold-start adaptation, shared-initialization adaptation, and frozen shared initialization, respectively, median server-side inference intervals were approximately 200, 201, and 200\,ms. Corresponding medians for latency from a TTS request to playback were 597, 648, and 639\,ms. Both summaries take the median across participants. The playback interval excludes LLM generation and any delay before the TTS request.

\begin{figure*}[!t]
  \centering
  \captionsetup{skip=4pt}
  \includegraphics[width=\linewidth,alt={Four study interfaces with Waiting above and Thinking, Answering, and the mobile companion below. Answer now activates the desktop agent, Stop answering interrupts it, and the circular mobile control supports push-to-talk input.}]{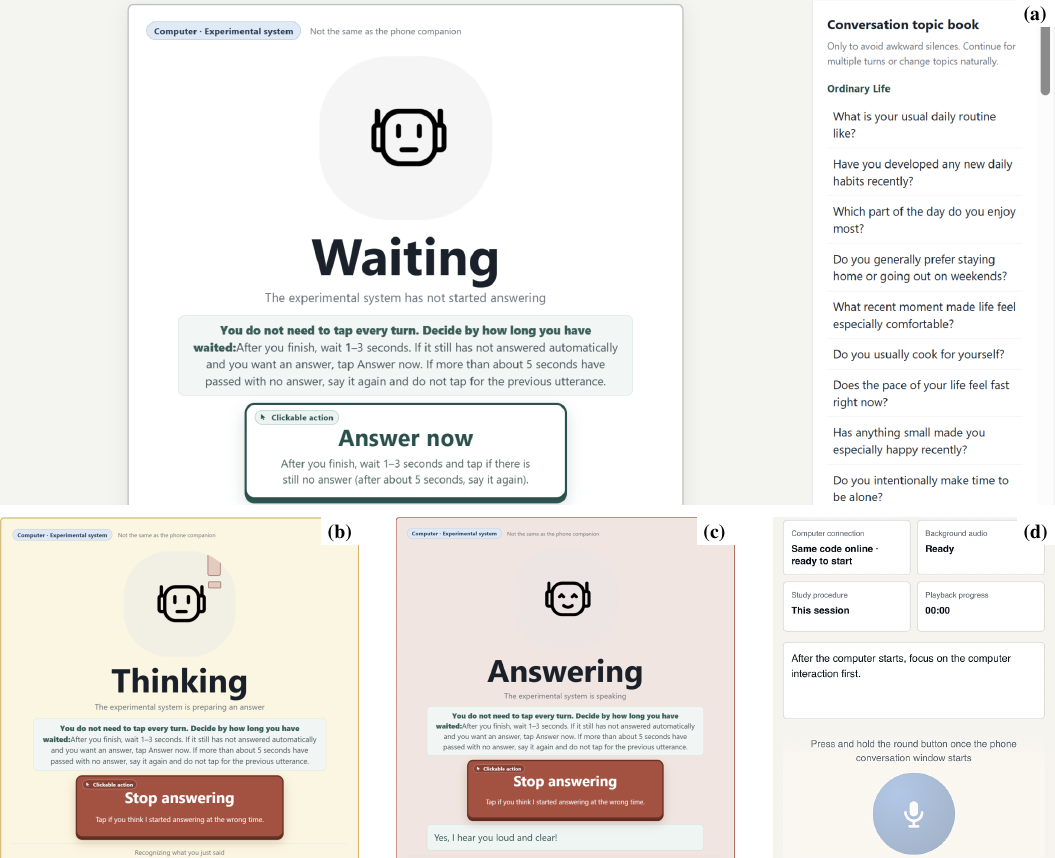}
  \caption{Study interfaces in English: (a) desktop Waiting with auxiliary topic prompts; (b) Thinking; (c) Speaking, displayed as Answering; (d) mobile companion. \emph{Answer now} activates and \emph{Stop answering} interrupts the desktop agent.}
  \Description{Four cropped interface panels in two rows. The upper panel shows the Waiting interface with its facial icon, an Answer now button, and conversation topics on the right. The lower-left Thinking and lower-middle Answering interfaces both contain a Stop answering button. The lower-right mobile companion shows connection and background-audio readiness and a circular hold-to-speak control.}
  \vspace{-4pt}
  \label{fig:study_interface}
\end{figure*}

Figure~\ref{fig:study_interface} shows the desktop states, the auxiliary list of conversation topics, and the mobile interface. At the same screen location, \emph{Answer now} enabled manual activation in Waiting, while \emph{Stop answering} enabled manual interruption in Thinking or Speaking, terminating generation or playback. State text and background colors distinguished system states, and Speaking appeared as \emph{Answering}. These actions retained the control and supervisory roles defined in Section~\ref{framework_overview}. Participants tapped when correction was needed, and a timer advanced the phases.

During continuous desktop conversations, the agent should respond after a completed question, request, or statement directed to it when a reply was expected, and wait during ongoing speech, brief pauses, or unyielded turns. Participants used \emph{Answer now} for missed responses and \emph{Stop answering} for premature ones.

We standardized correction opportunities with common waiting instructions. Participants were asked to wait 1 to 3\,s after finishing before requesting an unanswered reply. After about 5\,s without a reply, they were asked to repeat the utterance.

The control at the same screen location can switch between activation and interruption as the state changes. To prevent a tap intended for the previous state from triggering the new control action, we mask taps for 500\,ms after each transition. A masked tap displays a rejection message and is discarded before control, labeling, or model updates. Automatic triggering required the Waiting state, previously observed speech, at least 1.5\,s since the preceding trigger, and available training audio. The baseline response threshold was \(\tau_0=0.70\).

The mobile interface in Figure~\ref{fig:study_interface}(d) displayed connection and background-audio readiness. Participants held its circular PTT button to speak and released it to end input and await a reply. The desktop microphone captured speech during the hold, sending transcriptions to the mobile client and the acoustic stream to the desktop timing model. Mobile PTT delimited input. Desktop taps corrected timing.

Throughout each six-minute active block, the desktop listened, inferred timing under its assigned condition, and accepted corrections. It should wait during mobile conversations. Participants selected \emph{Stop answering} if it intervened. Desktop conversations thus alternated response and waiting opportunities, whereas mobile conversations concentrated waiting opportunities after addressee changes. All three conditions shared the task sequence and mobile PTT mechanism, differing in desktop initialization and online updating.

Speech directed to the mobile companion constituted user speech not addressed to the desktop, while the mobile TTS output constituted external agent speech. Both introduced speech interference. The fixed environmental track added acoustic interference. The environmental track played during both conversation phases, paused during mobile PTT and inter-block breaks, and then resumed from the same position. PTT standardized input boundaries and response requests on the mobile client, while recognition results, reply content, and playback latency varied with the interaction.

\subsubsection{Measures}

The primary outcome, \emph{eligible correction rate}, was calculated per participant and six-minute active block as the sum of eligible desktop activations and interruptions divided by valid desktop decision opportunities, then averaged across participants within condition. Both numerator and denominator covered desktop and mobile phases. Mobile PTT only delimited companion input. We excluded cancellations, accidental taps, and taps during breaks, applying identical eligibility criteria in the frozen condition independently of model updating. We additionally report each participant's block counts of activations, interruptions, eligible corrections, and raw taps.

Time spent responding or waiting can change how many decisions the desktop evaluates. We report corrections both per evaluated decision and per active minute of the fixed interaction budget. A decision opportunity is an eligible desktop timing evaluation. Successive evaluations of the same ongoing utterance count separately. Repeated utterances fall outside these tap-based outcomes. The event log has no dedicated repetition label, leaving their frequency unquantified.

Stage~1 reports raw taps per minute, calculated by dividing the total by 30 active minutes, and first-block correction rates in completion order. Tap frequency describes feedback intensity. Opening correction rates and parameter movement determine stopping. Stage~2 compares rates normalized by decision opportunities, supplemented by raw block counts to quantify burden per active minute.

The final questionnaire used 1--7 scales. Its measures based on the Subjective Assessment of Speech System Interfaces (SASSI)~\cite{hone2000towards} covered timing appropriateness, likeability, cognitive demand, annoyance, learnability, and efficiency. Additional constructs covered perceived control, predictability, correctability, understandability of change, and behavioral accommodation. Administered once after all five blocks, the questionnaire supported between-condition comparisons and retrospective ratings of perceived improvement and overall experience.

\FloatBarrier
\subsection{Stage 1: Shared Initialization}
\label{shared_initialization}
\paragraph{Design and Stopping Rule.}

To transfer earlier users' timing feedback to Stage~2 users, we used Stage~1 to establish a stable shared starting head. Sequential recruitment stopped when parameter and opening-correction plateaus met the prespecified confirmation rule. The first eight eligible participants established a comparable opening-behavior baseline. From participant~8 onward, each completion triggered these checks. Participant~1's head cloned the initial parameters \(\phi_0\). Each participant \(i>1\) started from \(\phi_{\rm sh}^{(i-1)}\), the mean final heads of preceding participants. All completed the same 30-minute online-learning session. With \(\phi_i^{\rm end}\) denoting participant~\(i\)'s complete final head, shared initialization after \(n\) completions was the equal-weight average
\[
\phi_{\rm sh}^{(n)}=\frac1n\sum_{i=1}^{n}\phi_i^{\rm end}.
\]
This implements model averaging sequentially across participants~\cite{mcmahan2017communication}. Each user-specific head used nominal step size \(\eta_{\rm head}\) with the feedback-count schedule in Section~\ref{implementation_details}, and shared initialization aggregated the completed participant's parameters with weight \(1/n\). Equal weighting applied to final decision heads.

On the parameter side, we measured adjacent relative movement of the complete vector of weights and bias
\[
d_\phi(n)=
\frac{\|\phi_{\rm sh}^{(n)}-\phi_{\rm sh}^{(n-1)}\|_2}
{\max(\|\phi_{\rm sh}^{(n-1)}\|_2,10^{-6})}.
\]
The denominator floor \(10^{-6}\) prevents division by zero. The parameter-plateau threshold was fixed at \(\tau_\phi=0.02\), a 2\% tolerance for relative movement between adjacent shared-initialization vectors. From \(n=8\), when the latest run of at least three participants had movements below this threshold, its first participant became the parameter-plateau candidate \(n_\phi\). The behavioral criterion used only each participant's first-block eligible correction rate \(q_i^{(1)}\), normalized by valid decision opportunities and excluding cancellations, accidental taps, and taps ineligible for training. Let \(Q_3(n)\) be the median opening correction rate of the latest three participants, with the first comparable baseline established at \(n=8\). The behavioral reference was the lowest median among these valid windows observed so far, \(Q^*(n)=\min_{8\le k\le n}Q_3(k)\), where \(k\) indexes each window's final participant. A window was on the behavioral plateau if its median fell at most \(\tau_q=0.002\) (0.2 percentage points) below or rose at most \(\tau_{\rm rebound}=0.02\) (2 percentage points) above the best preceding median.

Behavioral confirmation required at least five completions after the earliest window attaining the lowest median observed so far, enough for a non-overlapping window of three participants and two shifts by one participant. The latest window had to be on the behavioral plateau. The candidate \(n_q\) was the end index of the first window in the latest run of consecutive windows with the lowest median. The joint fitting point \(N^*\) was the later of \(n_q\) and \(n_\phi\). Only when sustained confirmation and the parameter plateau both held did the system lock the mean final heads of the first \(N^*\) participants as \(\phi_{\rm sh}^{*}=\phi_{\rm sh}^{(N^*)}\) and stop recruitment. Predetermined rebound criteria, a 22-participant confirmation cap, and a 50-participant hard cap provided additional stopping boundaries. Superscript \(*\) marks the running minimum in \(Q^*(n)\), the fitting point selected by the joint rule in \(N^*\), and the resulting locked shared head in \(\phi_{\rm sh}^{*}\).

\paragraph{Results.}

\begin{figure*}[t]
  \centering
  \includegraphics[width=\textwidth,alt={Two plots show opening correction rates and adjacent shared-initialization parameter movement across 20 sequential participants. Raw points determine stopping. Thick curves show Gaussian-kernel trends. Horizontal lines mark 1.06 percent and 0.02, respectively. Vertical lines mark the first 18 participants' fitting point.}]{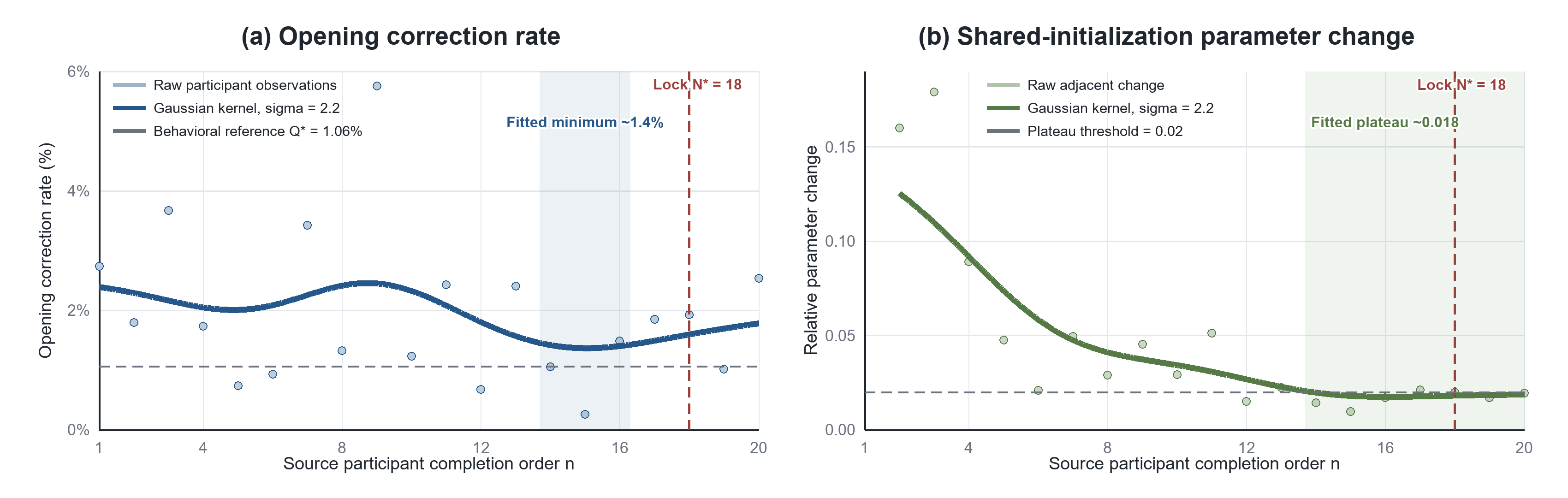}
  \caption{Stage~1 shared-head pretraining and saturation assessment. The left panel shows the opening correction rate, and the right shows adjacent relative movement of shared-initialization parameters. Faint points show raw participant observations. Thick curves show Gaussian-kernel trends with bandwidth 2.2. Horizontal dashed lines indicate \(Q^*=1.06\%\) and \(\tau_\phi=0.02\), respectively. The vertical lines mark the fitting point \(N^*=18\), confirmed after 20 completions. Stopping decisions used raw observations.}
  \Description{Two side-by-side plots with participant completion order from 1 to 20 on the horizontal axes. The left plot shows opening correction rates and the right adjacent relative parameter movement. Faint points show raw data. Thick curves show Gaussian-kernel smoothing with bandwidth 2.2. A vertical line in each plot marks participant 18.}
  \label{fig:share_fit}
\end{figure*}

The windows ending at participants~14--16 shared the lowest median opening correction rate of 1.06\% (rounded), giving \(n_q=14\). Five more completions confirmed the behavioral plateau at participant~19. The third consecutive parameter movement below 0.02 occurred at participant~20, giving \(n_\phi=18\) for the run of participants~18--20. Recruitment stopped and the mean of the first \(N^*=18\) final heads was locked as \(\phi_{\rm sh}^{*}=\phi_{\rm sh}^{(18)}\). Figure~\ref{fig:share_fit} shows the sequence. Table~\ref{tab:share_fit} provides participant-level values. Adjacent parameter movement declined from early values of 0.16--0.18 and fluctuated around 0.02 from approximately participant~12. Individual tap frequencies ranged from 0.50 to 5.17 per minute, with parameter movement and tap counts following different participant-level trajectories.

\setcounter{dbltopnumber}{2}

Stopping decisions used the raw observations shown in Figure~\ref{fig:share_fit}. Gaussian-kernel smoothing with a bandwidth of 2.2 participants summarized trends over completion order. Smoothed adjacent parameter movement reached 0.0174 at participant~16 and subsequently remained near 0.018. The raw movement at the fitting point \(n=18\) was 0.0198.

\begin{table}[t]
\centering
\caption{Stage~1 tap frequency, relative parameter movement, and opening correction rate for all 20 participants in completion order. Values of \(d_\phi\) use four decimal places. Stopping decisions used unrounded values.}
\label{tab:share_fit}
\fontsize{8}{9.4}\selectfont
\setlength{\tabcolsep}{2pt}
\begin{tabular*}{\linewidth}{@{\extracolsep{\fill}}rrrrp{0.28\linewidth}@{}}
\toprule
\shortstack{Order\\\(n\)} & Taps/min & \(d_\phi\) & \shortstack{Opening\\correction rate} & Decision \\
\midrule
1 & 1.27 & --- & 2.74\% & \shortstack[l]{Insufficient\\observations} \\
2 & 1.33 & 0.1602 & 1.80\% & \shortstack[l]{Insufficient\\observations} \\
3 & 2.60 & 0.1792 & 3.67\% & \shortstack[l]{Insufficient\\observations} \\
4 & 1.03 & 0.0891 & 1.73\% & \shortstack[l]{Insufficient\\observations} \\
5 & .93 & 0.0475 & .74\% & \shortstack[l]{Insufficient\\observations} \\
6 & .53 & 0.0209 & .93\% & \shortstack[l]{Insufficient\\observations} \\
7 & 3.37 & 0.0495 & 3.43\% & \shortstack[l]{Insufficient\\observations} \\
8 & .73 & 0.0289 & 1.33\% & First baseline \\
9 & 5.17 & 0.0455 & 5.76\% & Behavioral rebound \\
10 & 1.83 & 0.0292 & 1.23\% & Plateau observation \\
11 & 3.17 & 0.0512 & 2.43\% & Plateau observation \\
12 & .57 & 0.0150 & .67\% & Plateau observation \\
13 & 1.30 & 0.0223 & 2.40\% & Plateau observation \\
14 & .70 & 0.0143 & 1.06\% & Behavioral candidate \\
15 & .50 & 0.0096 & .26\% & Plateau observation \\
16 & 1.13 & 0.0170 & 1.49\% & Plateau observation \\
17 & 1.50 & 0.0212 & 1.85\% & Plateau observation \\
\textbf{18} & \textbf{1.60} & \textbf{0.0198} & \textbf{1.93\%} & \textbf{Fitting point} \\
19 & 1.00 & 0.0169 & 1.01\% & \shortstack[l]{Behavioral\\confirmation} \\
20 & 1.93 & 0.0195 & 2.53\% & Automatic lock \\
\bottomrule
\end{tabular*}
\end{table}

All 20 participants completed the final questionnaire. Aggregate means were 5.75 for correctability, 5.13 for perceived control, 5.10 for understandability of change, 5.08 for learnability, 5.00 for low annoyance, 4.93 for low cognitive demand, 4.75 for timing appropriateness, 4.75 for interaction efficiency, 4.70 for likeability, 4.40 for predictability, and 3.85 for low behavioral accommodation. Measured once at session end, these ratings reflect retrospective whole-session experience.

\subsection{Stage 2: Comparative Evaluation}
\label{extended_exp}

\subsubsection{Design and Participants}

The analysis included 48 new users, 16 per condition, who had not contributed to shared initialization. Sequential allocation prioritized the condition with the fewest completed study sessions among those still recruiting. Ties were resolved in the fixed order cold-start adaptation, shared-initialization adaptation, then frozen shared initialization. This was a deterministic balancing rule. Comparing the two online-adaptation conditions tests the benefit of shared initialization while allowing subsequent updating in both. Comparing shared-initialization adaptation with freezing tests the full adaptation procedure from a common starting point. The \emph{cold-start online adaptation} condition (\(C\)) started from the base head \(\phi_0\) (Section~\ref{implementation_details}), while \emph{shared-initialization online adaptation} (\(P\)) started from the locked shared head \(\phi_{\rm sh}^{*}\) (Section~\ref{shared_initialization}). Both updated the individual decision head, calibration bias, and history-conditioning parameters within the session, using \(\phi_0\) as the proximal reference. The \emph{frozen after shared initialization} condition (\(F\)) also started from \(\phi_{\rm sh}^{*}\), but held the user-specific head, calibration bias, and history-conditioning parameters fixed throughout. Taps were recorded without entering the adaptation history. History-dependent threshold shifts and logit residuals remained zero.

Participants were predominantly 18--37, including 25 women, 22 men, and one who declined to disclose gender. Seventeen were students, 30 non-students, and one had unclassified occupation. Twenty-four reported pronunciation close to the system language's standard variety, and 19 reported a mild regional accent. None reported a hearing or speech condition.

\begin{table}[t]
\centering
\caption{Stage~2 mean eligible correction rates per decision opportunity, by six-minute block. \(C\), \(P\), and \(F\) denote cold-start adaptation, shared-initialization adaptation, and frozen shared initialization, respectively.}
\label{tab:block_rates}
\small
\begin{tabular*}{\linewidth}{@{\extracolsep{\fill}}lrrr@{}}
\toprule
Six-minute block & \(C\) & \(P\) & \(F\) \\
\midrule
1 & 1.61\% & 0.88\% & 1.32\% \\
2 & 1.35\% & 0.83\% & 0.80\% \\
3 & 0.91\% & 0.69\% & 0.90\% \\
4 & 1.01\% & 0.72\% & 0.82\% \\
5 & 0.72\% & 0.65\% & 0.67\% \\
\bottomrule
\end{tabular*}
\end{table}

\begin{table*}[t]
\centering
\begin{minipage}[t]{\columnwidth}
\centering
\caption{Poisson GEE incidence rate ratios (IRR) for eligible corrections per active minute, with 95\% CIs. \(C\), \(P\), and \(F\) denote cold-start adaptation, shared-initialization adaptation, and frozen shared initialization, respectively.}
\label{tab:participant_count_model}
\small
\setlength{\tabcolsep}{3pt}
\begin{tabular*}{\linewidth}{@{\extracolsep{\fill}}lrrr@{}}
\toprule
Eligible-correction contrast & IRR & 95\% CI & \(p\) \\
\midrule
Block 1, \(C\) / \(P\) & 1.69 & [1.15, 2.49] & .008 \\
Block 1, \(F\) / \(P\) & 1.67 & [1.03, 2.71] & .037 \\
Whole session, \(C\) / \(P\) & 1.59 & [1.09, 2.32] & .015 \\
Whole session, \(F\) / \(P\) & 1.53 & [0.98, 2.40] & .061 \\
\(C\), block 5 / block 1 & 0.51 & [0.38, 0.68] & \(<.001\) \\
\(P\), block 5 / block 1 & 0.75 & [0.49, 1.14] & .176 \\
\(F\), block 5 / block 1 & 0.54 & [0.39, 0.76] & \(<.001\) \\
\bottomrule
\end{tabular*}
\end{minipage}\hfill
\begin{minipage}[t]{\columnwidth}
\centering
\caption{Correction-rate differences relative to shared-initialization adaptation, in percentage points (pp), with 95\% CIs. Condition labels follow Table~\ref{tab:participant_count_model}.}
\label{tab:observable_arm_contrasts}
\small
\setlength{\tabcolsep}{3pt}
\begin{tabular*}{\linewidth}{@{\extracolsep{\fill}}lrr@{}}
\toprule
Block & \(C-P\) (pp), 95\% CI & \(F-P\) (pp), 95\% CI \\
\midrule
1 & 0.73 [0.25,1.20] & 0.44 [-0.18,1.09] \\
2 & 0.52 [-0.19,1.24] & -0.03 [-0.61,0.52] \\
3 & 0.22 [-0.44,0.81] & 0.20 [-0.46,0.75] \\
4 & 0.29 [-0.21,0.80] & 0.09 [-0.35,0.49] \\
5 & 0.06 [-0.31,0.40] & 0.02 [-0.35,0.37] \\
Whole session & 0.36 [-0.11,0.80] & 0.15 [-0.33,0.58] \\
\bottomrule
\end{tabular*}
\end{minipage}
\end{table*}

\begin{table}[t]
\centering
\caption{Within-session declines under frozen shared initialization and differences in decline (shared-initialization adaptation minus frozen shared initialization), in percentage points (pp), with 95\% CIs. Early and late periods cover blocks~1--3 and~3--5, respectively.}
\label{tab:observable_change_contrasts}
\fontsize{8}{9.4}\selectfont
\setlength{\tabcolsep}{2pt}
\begin{tabular*}{\linewidth}{@{\extracolsep{\fill}}lrr@{}}
\toprule
Interval & \shortstack{Frozen decline\\{}(pp), 95\% CI} & \shortstack{Difference in decline\\{}(pp), 95\% CI} \\
\midrule
1\(\rightarrow\)2 & 0.52 [0.16,0.94] & -0.47 [-0.99,-0.01] \\
2\(\rightarrow\)3 & -0.10 [-0.32,0.13] & 0.23 [-0.08,0.54] \\
3\(\rightarrow\)4 & 0.08 [-0.21,0.37] & -0.11 [-0.47,0.28] \\
4\(\rightarrow\)5 & 0.14 [-0.08,0.36] & -0.08 [-0.40,0.24] \\
1\(\rightarrow\)3 cumulative & 0.43 [0.12,0.76] & -0.24 [-0.74,0.21] \\
1\(\rightarrow\)4 cumulative & 0.51 [0.08,0.99] & -0.35 [-0.86,0.12] \\
1\(\rightarrow\)5 whole session & 0.65 [0.18,1.19] & -0.42 [-1.03,0.13] \\
Early\(\rightarrow\)late & 0.21 [0.01,0.44] & -0.10 [-0.35,0.13] \\
\bottomrule
\end{tabular*}
\end{table}

\begin{figure*}[!t]
  \centering
  \includegraphics[width=\textwidth,alt={Two plots show mean activation and interruption taps over five blocks in three conditions. Cold-start online adaptation has the most opening interruptions. Frozen after shared initialization has the most opening activations. Shared-initialization online adaptation generally requires fewer taps. Points show raw means. Curves show moving-average trends.}]{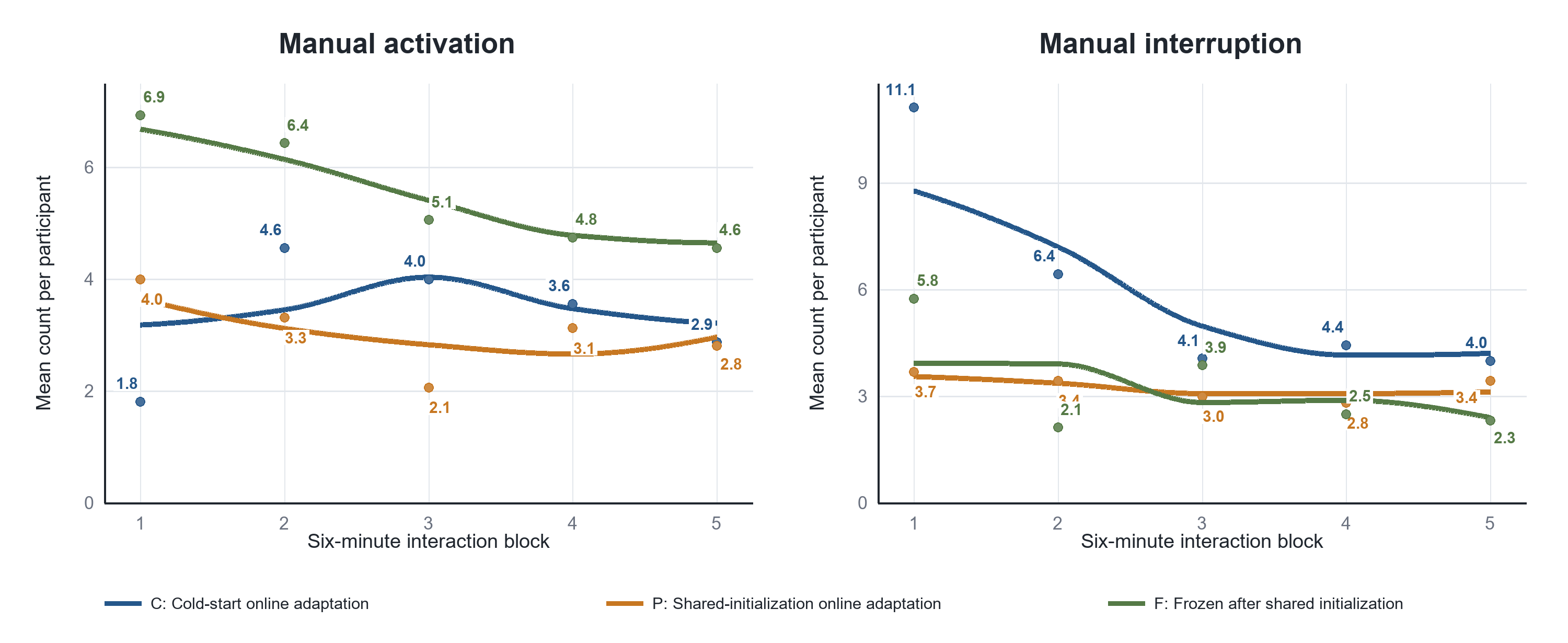}
  \caption{Stage~2 mean manual activation and interruption taps per participant in five six-minute blocks. Point labels give raw block means. Thick curves use shape-preserving cubic interpolation of three-point moving averages, using two adjacent points at the endpoints. Each condition has \(n=16\).}
  \Description{Two plots show activation and interruption counts. Faint points show the five raw block means, and thick curves interpolate three-point moving averages. Cold-start online adaptation has the most opening interruptions, frozen after shared initialization the most opening activations, and shared-initialization online adaptation generally lower counts of both.}
  \label{fig:user_exp}
\end{figure*}

\begin{figure*}[t]
  \centering
  \includegraphics[width=\textwidth,alt={Correction rates over five blocks for cold-start online adaptation, shared-initialization online adaptation, and frozen after shared initialization. Points show raw means, error bars participant-level 95 percent bootstrap intervals, and dashed curves moving-average trends. Shared-initialization online adaptation has the lowest opening mean. Later means converge.}]{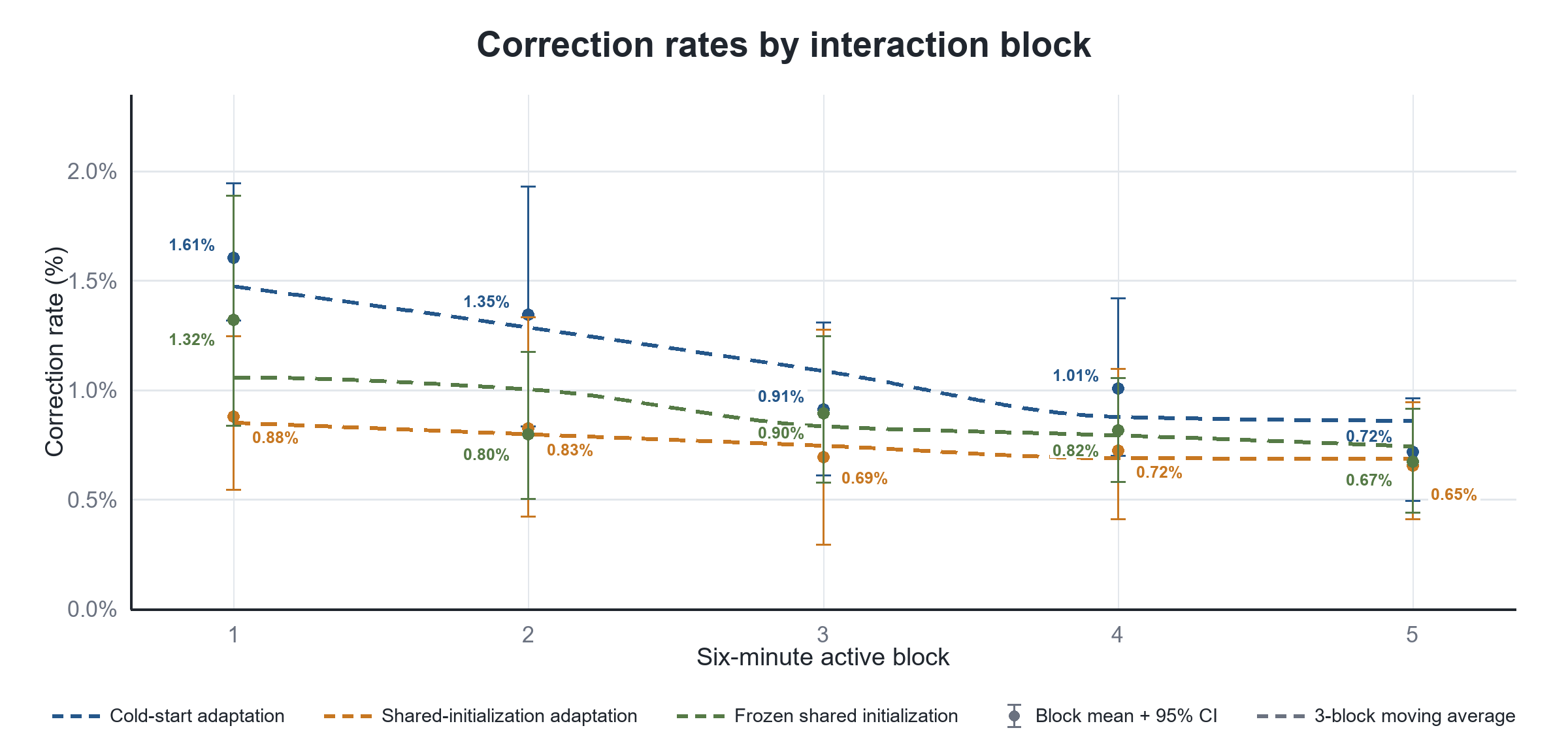}
  \caption{Raw block correction rates and temporal trends for the three conditions, with 16 participants each. Point labels give raw block means. Error bars are pointwise 95\% bootstrap intervals resampling participants' complete five-block records. Dashed curves use shape-preserving cubic interpolation of three-point moving averages to summarize trends, using two adjacent points at the endpoints.}
  \Description{Points and error bars show five raw block means and participant-level intervals for cold-start online adaptation, shared-initialization online adaptation, and frozen after shared initialization. Matching dashed curves interpolate three-point moving averages. Shared-initialization online adaptation has the lowest opening mean, and the conditions become closer in later blocks.}
  \label{fig:extended_core}
\end{figure*}
\subsubsection{Analysis Procedures}

Correction-rate contrasts used participant-level nonparametric bootstrap intervals~\cite{efron1992bootstrap}. From the raw block correction rates underlying Table~\ref{tab:block_rates}, we calculated differences from shared-initialization adaptation for cold-start adaptation and frozen shared initialization, subtracting the reference rate from each comparison rate. Within each condition, we resampled 16 participants' complete five-block records with replacement, using 20,000 replicates and a fixed random seed. Table~\ref{tab:observable_arm_contrasts} reports percentile 95\% bootstrap intervals. Whole-session estimates weight blocks equally.

Within-session changes comprise the frozen condition's decline and the difference in decline (shared-initialization online adaptation minus frozen shared initialization). Early and late periods equally weight blocks~1--3 and~3--5, respectively, sharing block~3. Table~\ref{tab:observable_change_contrasts} uses the same participant-level resampling and preserves estimate signs.

Stage~2 comprised 240 participant--block observations from 48 participants with five blocks each. A supplementary analysis using Poisson generalized estimating equations (GEE)~\cite{liang1986longitudinal} modeled eligible correction counts \(Y_{ib}\) for participant \(i\) in block \(b\) with a log link, categorical condition and block main effects and their interaction, and offset \(\log T_{ib}\) with coefficient fixed at 1. Effective duration \(T_{ib}\), measured in minutes, excludes breaks and unanalyzable segments. Completion criteria ensured \(T_{ib}=6.00\) active minutes per block, totaling 1,440 participant-minutes. Models clustered by participant, with exchangeable working correlation and robust sandwich standard errors. Block-specific and within-condition contrasts used the condition\(\times\)block model. Whole-session contrasts used only condition and block main effects. We report incidence rate ratios per active minute, \(\mathrm{IRR}=\exp(\Delta\beta)\), and 95\% CIs, where \(\Delta\beta\) is the corresponding log-rate coefficient contrast. An IRR above one indicates more corrections per active minute in the numerator condition or block. Eligible-correction estimates are reported in Table~\ref{tab:participant_count_model}. The same specification separately analyzed raw activation and interruption counts as secondary outcomes.

The available cohort comprises 16 participants per condition. Precision is assessed through participant-level confidence intervals~\cite{lakens2022sample}. Exploratory sensitivity analyses use the bias-reduced GEE sandwich covariance implemented in statsmodels, leave-one-participant-out refits, and baseline-covariate adjustment. Reported \(p\)-values are two-sided and nominal. Effect sizes and intervals guide interpretation. All sensitivity analyses retain the participant as the sampling unit and the original condition, block, interaction, and active-duration specifications. Accent is coded as near-standard versus other pronunciation, familiarity uses the baseline rating, and age is coded as 18--27 versus at least 28. To limit the number of correlated predictors, each extended specification adds one demographic variable to accent and familiarity. Complete-case gender and occupation analyses each contain 47 participants. Their corresponding unadjusted subsets separate changes in sample composition from adjustment.

The final questionnaire included seven open-ended questions and conditional follow-ups. All 48 participants answered questions about overall impressions, unexpected events, corrective control, user adaptation, companion effects, and desired improvements; 47 described the most noticeable change. The 30 participants who rated timing improvement at least 6 received a follow-up asking for evidence. The first author coded responses by question. Each participant could receive multiple codes. Frequencies describe self-reported experiences in this sample.
\begin{table*}[t]
\centering
\caption{Exploratory GEE sensitivity estimates reported as IRR [95\% CI] with bias-reduced covariance. A plus sign adds the indicated variable to accent and familiarity. \(C\), \(P\), and \(F\) denote cold-start adaptation, shared-initialization adaptation, and frozen shared initialization, respectively.}
\label{tab:sensitivity}
\footnotesize
\setlength{\tabcolsep}{3pt}
\begin{tabular*}{\textwidth}{@{\extracolsep{\fill}}lrrrrr@{}}
\toprule
 & & \multicolumn{2}{c}{Opening block} & \multicolumn{2}{c}{Whole session} \\
Covariates & \(n\) & \(C\) relative to \(P\) & \(F\) relative to \(P\) & \(C\) relative to \(P\) & \(F\) relative to \(P\) \\
\midrule
Unadjusted & 48 & 1.69 [1.12, 2.55] & 1.67 [1.00, 2.80] & 1.59 [1.07, 2.38] & 1.53 [0.95, 2.47] \\
Accent, familiarity & 48 & 1.59 [1.03, 2.44] & 1.53 [0.92, 2.53] & 1.49 [0.96, 2.31] & 1.40 [0.86, 2.27] \\
+ gender & 47 & 1.49 [0.95, 2.34] & 1.50 [0.89, 2.54] & 1.39 [0.88, 2.19] & 1.36 [0.83, 2.23] \\
+ student status & 47 & 1.43 [0.99, 2.05] & 1.55 [0.99, 2.43] & 1.33 [0.91, 1.93] & 1.41 [0.92, 2.16] \\
+ age group & 48 & 1.44 [0.99, 2.09] & 1.72 [1.12, 2.63] & 1.33 [0.90, 1.96] & 1.55 [1.04, 2.32] \\
Unadjusted gender subset & 47 & 1.69 [1.12, 2.55] & 1.72 [1.02, 2.90] & 1.60 [1.07, 2.38] & 1.58 [0.97, 2.57] \\
Unadjusted occupation subset & 47 & 1.58 [1.07, 2.34] & 1.57 [0.95, 2.58] & 1.49 [1.02, 2.18] & 1.44 [0.91, 2.28] \\
\bottomrule
\end{tabular*}
\end{table*}

\subsubsection{Results}

\paragraph{RQ1. How does shared initialization affect the cold-start burden?}
Table~\ref{tab:block_rates} reports participant-averaged correction rates. The opening rate was lowest in shared-initialization adaptation at 0.88\%, versus 1.61\% in cold-start adaptation and 1.32\% in frozen shared initialization. Final-block rates converged to 0.72\%, 0.65\%, and 0.67\% in cold-start adaptation, shared-initialization adaptation, and frozen shared initialization, respectively. In that order, whole-session means per participant were 44.2, 29.8, and 43.6 eligible corrections; 46.9, 31.7, and 44.3 raw taps; and approximately 4417, 4620, and 5227 opportunities. First-three-block correction rates averaged 1.24\%, 0.80\%, and 0.99\%, versus 0.85\%, 0.69\%, and 0.77\% over the last three. Tables report raw block means. Moving averages visualize trends.

Figure~\ref{fig:user_exp} further decomposes corrections. In block~1, mean manual interruption counts per participant were 11.1, 3.7, and 5.8 in cold-start adaptation, shared-initialization adaptation, and frozen shared initialization, respectively. Mean manual activation counts were 1.8, 4.0, and 6.9 in the same order. Across blocks 1--5, mean manual activation counts in cold-start adaptation were 1.8, 4.6, 4.0, 3.6, and 2.9. Fifteen of its 16 participants first corrected the system through interruption. In block~1, cold-start adaptation relative to shared-initialization adaptation had an eligible-correction \(\mathrm{IRR}=1.69\) with 95\% CI [1.15, 2.49], a raw-interruption \(\mathrm{IRR}=3.02\) [1.91, 4.76], and a raw-activation \(\mathrm{IRR}=0.45\) [0.25, 0.81]. For frozen shared initialization relative to shared-initialization adaptation, the block~1 eligible-correction estimate was \(\mathrm{IRR}=1.67\) [1.03, 2.71], with a raw-activation \(\mathrm{IRR}=1.73\) [1.11, 2.71]. By block~5, the eligible-correction contrast for cold-start relative to shared-initialization adaptation was \(\mathrm{IRR}=1.15\) [0.71, 1.84], with the interval including the null value.

Cold-start adaptation exceeded shared-initialization adaptation by 0.73 percentage points [0.25,1.20] in block~1 and 0.06 [-0.31,0.40] in block~5 (Table~\ref{tab:observable_arm_contrasts}).

\paragraph{RQ2. User and system adaptation.}
The system updated individual decision heads in response to 676 of 708 eligible corrections in cold-start adaptation and 444 of 476 in shared-initialization adaptation. Calibration bias and history parameters also updated during head warm-up. Frozen shared initialization recorded 698 eligible corrections and zero parameter updates.

Mean interruption counts over the first three blocks were 7.2, 3.4, and 3.9 in cold-start adaptation, shared-initialization adaptation, and frozen shared initialization, versus 4.2, 3.1, and 2.9 over the last three. Whole-session per-block means were 6.0, 3.3, and 3.3. Activation counts averaged 3.5, 3.1, and 6.2 over the first three blocks, versus 3.5, 2.7, and 4.8 over the last three. In the count model, the IRRs for eligible corrections in block~5 relative to block~1 were 0.51 [0.38, 0.68] for cold-start adaptation, 0.75 [0.49, 1.14] for shared-initialization adaptation, and 0.54 [0.39, 0.76] for frozen shared initialization.

The whole-session eligible-correction IRR for frozen relative to shared-initialization adaptation was 1.53 [0.98, 2.40] (\(p=.061\)). The ratio of the IRR for block~5 relative to block~1 in shared-initialization adaptation to the corresponding IRR in frozen shared initialization was 1.38 [0.80, 2.36] (\(p=.244\)). Both intervals include one.

The frozen condition declined by 0.65 percentage points [0.18,1.19] over the session. The difference in decline (shared-initialization adaptation minus frozen shared initialization) was -0.42 [-1.03,0.13], as reported in Table~\ref{tab:observable_change_contrasts}.

Of 48 participants, 31 reported slower speech, longer pauses, or waiting 1 to 3 seconds first. Eight reported no change in speaking style. Figure~\ref{fig:extended_core} shows the block-level correction trajectories.

\paragraph{Sensitivity of condition comparisons.}
Table~\ref{tab:sensitivity} reports exploratory covariate-adjusted and bias-reduced estimates.
The opening estimate for cold-start relative to shared-initialization adaptation remained above one under bias-reduced covariance, at 1.69 [1.12, 2.55] with \(p=.013\). Adjusting for accent and speech-technology familiarity gave 1.59 [1.03, 2.44], as reported in Table~\ref{tab:sensitivity}.

Across leave-one-participant-out whole-session refits using the original robust covariance, IRRs for cold-start relative to shared-initialization adaptation ranged from 1.49 to 1.75 with \(p=.006\)--.041, and IRRs for frozen relative to shared-initialization adaptation from 1.39 to 1.69 with \(p=.029\)--.167.

\paragraph{Perceived improvement, overall experience, and construct ratings.}
Table~\ref{tab:usability_results} reports 1--7 means, emphasizing the direction and magnitude of experiential differences. Across conditions, perceived-improvement means ranged from 5.44 to 6.00 and overall-experience means from 4.81 to 5.88. Shared-initialization online adaptation had the highest overall score for low cognitive demand, at 5.50. All groups' overall ratings fell in the middle and upper parts of the scale. In cold-start adaptation, shared-initialization adaptation, and frozen shared initialization, respectively, predictability means were 4.90, 5.17, and 5.04; perceived control, 5.94, 5.38, and 5.38; correctability, 5.88, 5.75, and 5.75; understandability of change, 5.38, 5.56, and 4.81; and low behavioral accommodation, 4.31, 4.56, and 4.31.

\begin{table}[t]
\centering
\caption{Stage~2 mean ratings (1--7) of perceived improvement and overall experience, collected once at session end. \(C\), \(P\), and \(F\) denote cold-start adaptation, shared-initialization adaptation, and frozen shared initialization, respectively.}
\label{tab:usability_results}
\small
\setlength{\tabcolsep}{4pt}
\tagpdfsetup{table/header-rows={1,2,10,11}}
\begin{tabular*}{\linewidth}{@{\extracolsep{\fill}}lrrr@{}}
\toprule
\multicolumn{4}{c}{Perceived improvement over interaction} \\
Dimension & \(C\) & \(P\) & \(F\) \\
\midrule
Timing appropriateness & 5.81 & 5.63 & 5.56 \\ Likeability & 5.63 & 5.69 & 5.50 \\
Low cognitive demand & 5.75 & 6.00 & 5.94 \\ Low annoyance & 5.50 & 5.44 & 5.63 \\
Learnability & 5.88 & 5.88 & 5.75 \\ Perceived control & 5.81 & 5.56 & 5.81 \\
Interaction efficiency & 5.63 & 5.63 & 5.75 \\ \addlinespace \multicolumn{4}{c}{Overall experience} \\
Dimension & \(C\) & \(P\) & \(F\) \\ \midrule
Timing appropriateness & 5.63 & 5.69 & 5.25 \\ Likeability & 5.63 & 5.88 & 5.69 \\
Low cognitive demand & 4.81 & 5.50 & 4.94 \\ Low annoyance & 5.25 & 5.31 & 5.25 \\
Learnability & 5.44 & 5.63 & 5.31 \\ Interaction efficiency & 5.38 & 5.50 & 5.50 \\ \bottomrule
\end{tabular*}
\end{table}

\paragraph{Open-ended questionnaire responses.}
Approximately one-third spontaneously described appropriate timing or increasingly smooth interaction. Others commonly mentioned short replies, mechanical voices, memory, or everyday topics. Unexpected events primarily involved recognition of regional speech or place names, environmental noise, desktop interventions during mobile conversations, responses before users had finished speaking, and agent-initiated follow-up questions.

Twenty-five of 48 explicitly reported that tapping started or stopped the agent as expected and restored control. Four reported needing repeated taps or finding a tap ineffective at that moment. Approximately 40 mentioned distraction, missed speech, or interruptions due to background sounds, construction noise, or mobile conversations. Approximately 14 described little impact. These codes were not mutually exclusive. About one-quarter of the responses about the most noticeable change concerned more timely responses, fewer interruptions, or better pacing. Among the 30 high-improvement respondents who received the conditional follow-up, 15 mentioned increasingly accurate or faster timing or fewer interruptions. This frequency characterizes how high-rating respondents explained improvement. Suggestions emphasized regional-speech and noise recognition, fewer premature responses, somewhat longer replies, and more natural voices.

\paragraph{Pre-study characteristics and experiential associations.}
Age, occupation, accent, and everyday language use strongly co-occurred. All 17 students were 18--27, while all 22 participants aged at least 28 were non-students. Twenty-two reported pronunciation close to the standard variety and everyday use of the dialogue system's selected language; 17 reported an accent and everyday regional-dialect use. Of the 24 participants with an accent, 19 described it as mild. Descriptive within-sample associations included fewer perceived unexpected response initiations among those aged at least 28, with scores of 5.50 versus 4.08. Overall likeability was higher among participants with accents, at 6.17 versus 5.29, or primarily using a regional dialect, at 6.32 versus 5.34. Non-students more often described improved timing as the most noticeable change, with 13/30 versus 0/17. Participants more familiar with speech technology more often reported deliberately changing their wording, with 15/27 versus 2/21, or the system gradually keeping pace, with 12/27 versus 1/21.

\subsection{Data Resource and Governance}

The study records speech-derived features, decision opportunities, corrections, system states, response latency, model outputs, online updates, and quality telemetry. Event timestamps, block, phase, and dialogue-source labels preserve interaction context within a hierarchy of participants, sessions, blocks, and events. Breaks contain only boundary and quality-audit records. The behavioral analyses use event records from valid active blocks.

Participants provided informed consent for study participation. Data collection, use, and retention followed each participant's documented consent choices. Compensation identities were stored separately from research records, which exclude names, contact information, and reimbursement identities.

To support model and algorithm evaluation and selection for response initiation timing, we provide 1,000 labeled pilot recordings. The corpus contains human-recorded 15-second clips with scenario-intention labels, collection order, and pooled acoustic features. Its documentation records file hashes and annotation provenance following data-statement principles~\cite{bender2018data}. The data owner authorized public distribution. The recordings retain voice characteristics and spoken content.

Participant and session pseudonyms use HMAC-SHA256~\citep{turner2008keyed,dang2015secure}. Identifiers, device metadata, and recruitment mappings are stored separately. Individual trajectories, linkage keys, and participant-specific weights remain private.

\section{Discussion}
\label{discussion}
We interpret the technical evaluation of BPRHT and the user study of the Tap-to-Adapt framework to understand how corrective feedback supports personalized response timing. We relate shared initialization and adaptation between users and the system to correction burden and reported experience, then discuss limitations and future research.

\subsection{Interpretation of the Findings}

\subsubsection{RQ1. Shared Initialization and Opening Correction Burden}

Shared initialization's clearest observed benefit was reducing the interventions needed at the beginning of interaction. The gap between cold-start and shared-initialization adaptation narrowed substantially by the final block (Tables~\ref{tab:block_rates} and~\ref{tab:observable_arm_contrasts}). This temporal pattern suggests that prior interaction experience is particularly useful while little personal feedback is available. The opening point estimates for cold-start relative to shared-initialization adaptation pointed in the same direction across the sensitivity specifications (Table~\ref{tab:sensitivity}).

Cold-start online adaptation had the fewest opening activations but the most opening interruptions (Figure~\ref{fig:user_exp}). Compared with shared-initialization online adaptation, cold-start online adaptation combined fewer activation requests with a higher interruption burden and overall correction rate. Cold-start online adaptation may initially respond too readily, including when users would prefer continued waiting, before enough personal feedback has accumulated to align response timing with their expectations. Such a tendency could reduce manual activation requests while increasing the number of responses users wish to interrupt. Shared-initialization online adaptation had fewer activations and interruptions than frozen shared initialization at the opening. By the final block, however, shared-initialization online adaptation still had fewer activations, while frozen shared initialization had fewer interruptions. The relative advantage thus depended on the feedback channel and stage of interaction.

The Stage~1 trajectories show how a useful shared starting point can coexist with variation between users. Parameter movement became smaller while opening correction rates continued to fluctuate (Figure~\ref{fig:share_fit} and Table~\ref{tab:share_fit}). Their different trajectories suggest that shared-head stability summarizes accumulated feedback across users with differing correction patterns. The resulting starting point therefore leaves room for subsequent personal adjustment.

\subsubsection{RQ2. User and System Adaptation}

The quantitative trajectories and coded responses suggest adaptation on both sides of the interaction. Corrections declined both where parameter updates were logged and under freezing (Figure~\ref{fig:extended_core}). The speaking-and-waiting adjustments reported by 31 participants provide a behavioral account of how users could contribute. P16 in shared-initialization adaptation wrote, ``Yes, I spoke much more slowly than usual, paused frequently, and articulated much more clearly.'' Such adjustments may make acoustic cues at speech boundaries clearer, helping coordinate the interaction through changes in user behavior.
\par The cold-start activation trajectory suggests a possible feedback sequence. Opening corrections were dominated by interruptions, followed by a rise and then a fall in activations (Figure~\ref{fig:user_exp}). Repeated early interruptions could temporarily shift decisions toward waiting, increasing the need for activation; subsequent positive feedback could then rebalance initiation. This interpretation connects the two observed feedback trajectories with the correction directions illustrated in Section~\ref{sec:feedback_example}.
\par Participants also described changes in the system\textquotesingle{}s behavior. P41 in cold-start adaptation observed, ``The desktop assistant gradually stopped answering at the same time as the phone assistant\textquotesingle{}s notification sounds.'' This account describes perceived improvement around a recurring environmental cue. Together with the timing-improvement codes in Section~\ref{extended_exp}, it adds experiential meaning to the aggregate declines: users described both adjusting their own speech and experiencing more appropriate responses. These complementary accounts support interpreting alignment as a developing interaction between user practices and system behavior.

\subsubsection{Perceived Control and Accommodation}
The control-related codes help explain why direct intervention remained valuable alongside automatic timing. Twenty-five participants reported that tapping restored control. P18 in frozen shared initialization wrote, ``When noise came from the phone, the desktop AI agent would recognize it incorrectly, and I would press Stop answering. Yes, I had control over the interaction.'' This account from the frozen condition illustrates how immediate intervention can support perceived control. This helps interpret the favorable correctability and control ratings in both stages (Section~\ref{shared_initialization} and Table~\ref{tab:usability_results}): users could regain initiative when automatic responses did not match their intentions.

The accommodation codes place these favorable experiences in context. Slower speech and longer pauses may improve coordination and may also involve additional user effort. The accommodation ratings and reports of altered speaking practices therefore suggest that fewer corrections and perceived improvement can coexist with such effort. The timing-improvement codes emphasize more timely replies, fewer interruptions, and better pacing, while the control codes emphasize the ability to intervene. Taken together, these findings favor evaluating timing adaptation through expressed corrections, perceived control, and the speaking-and-waiting adjustments users report.

\subsection{Limitations}

\paragraph{Study Design and Generalizability.}
The study examined within-session adaptation over 30 active minutes. Stage~1 followed one participant sequence. Stage~2 balanced completed-session counts with deterministic tie-breaking, with 16 participants per condition. In Stage~1, parameter movement and opening correction rates fluctuated before the plateau (Figure~\ref{fig:share_fit} and Table~\ref{tab:share_fit}), and each additional head received a decreasing averaging weight. The confirmed participant count therefore reflects this sequence, the feedback accumulated within the fixed interaction duration, and the chosen plateau criteria. Different sequences or durations could change the amount and composition of feedback and the stopping point. Establishing a task-wide participant requirement would require replication across such conditions. Sessions took place primarily in a university setting, and allocation balanced completed-session counts rather than demographic characteristics. Nonrandom allocation leaves condition comparisons susceptible to recruitment and participant-composition effects. The single-session design limits estimates of long-term stability. Session-end reports document speaking and waiting adjustments at the session scale. Conversation topics could change freely, so the frozen trajectory does not isolate user accommodation from changes in the conversation. Testing the proposed feedback-driven account of the activation trajectory requires linking event-level corrections to subsequent predictions under comparable response opportunities. The condition comparisons estimate combined effects rather than independent proportional contributions of shared initialization, system updating, and user adjustment; a condition frozen at base initialization could extend these comparisons.
\paragraph{Sensitivity of Condition Comparisons.} For cold-start relative to shared-initialization adaptation, covariate adjustment shifted the rate-ratio estimates toward one (Table~\ref{tab:sensitivity}). The opening interval remained above one with accent and familiarity adjustment, while the additional demographic specifications included one; all covariate-adjusted whole-session intervals for this contrast also included one. For frozen relative to shared-initialization adaptation, most intervals included one, while age-adjusted opening and whole-session intervals were above one. This variation makes the evidence for an additional benefit of online adaptation sensitive to the chosen specification. Adjusted point estimates also differed numerically from unadjusted estimates for the same complete-case subsets, showing sensitivity to joint adjustment at a fixed sample composition. Intervals spanning one leave the magnitude and direction of the underlying rate contrast uncertain. The 95\% interval for the difference in first-to-last decline between shared-initialization adaptation and freezing also included zero (Table~\ref{tab:observable_change_contrasts}), leaving the incremental behavioral contribution of online updating unresolved.
\paragraph{Feedback Semantics and Learning Scope.}
Participants reported recognition difficulties, short replies, and mechanical voices alongside timing-related experiences (Section~\ref{extended_exp}). Binary tap labels do not encode the participant's reason for intervening, so response-quality concerns cannot be separated from timing-related interventions using the label alone. Tap logs record expressed interventions rather than independent assessments of every response opportunity; missed or unwanted responses that elicited no tap are therefore outside what correction counts identify. These counts should be interpreted alongside reported changes in interaction practices. Content-level long-term memory may influence subsequent taps through changes in replies and user expectations. The study evaluates response initiation, with stopping retained as an immediate user control. The 1 to 3\,s waiting instruction and approximately 5\,s repetition rule standardized correction procedures and reduced procedural variation across conditions. These instructions constrain expressed timing preferences and differ from spontaneous interaction. Correction counts and favorable ratings may overlook accommodation costs such as slowing speech or lengthening pauses. Replayed samples remain statistically dependent across updates. Pilot labels and the first author's qualitative coding received no independent review. User-specific generalization requires independent later observations and cross-user model matching. Generalization to unrestricted waiting policies remains open.

\paragraph{Deployment Conditions and Privacy.}
LLM processing, synthesis, and playback contribute to audible response latency and can affect tap timing. Quality checks and event audits identify some failures. Residual latency remains part of the feedback loop. The turn-taking comparisons evaluated simplified implementations under a common binary-label protocol. The history-conditioning hyperparameters were set during implementation without independent development-set estimation. Selection used fixed synthetic stimuli, with gate and threshold settings that differed from continuous real-user deployment, as detailed in Section~\ref{sec:history_selection}. The five-seed check varied simulated block shifts and within-sentence pauses and included the selection seed, leaving possible selection optimism. These conditions limit direct extrapolation of simulated component effects. Added acoustic statistics yielded little calibration change and no decision-level gain on the fixed streams. Frozen DistilHuBERT representations may vary in robustness across accents, acoustic environments, and devices. Raw speech, transcripts, and individual models also carry re-identification risks.

\setcounter{dbltopnumber}{2}

\subsection{Future Work}

\paragraph{Cumulative Adaptation across Contexts.}
The present protocol retained learned personal parameters when interaction resumed after one-minute breaks. Future work could extend this continuity to the accumulation and transfer of corrective experience across settings and longer intervals between uses. For example, longitudinal studies could follow the same users between quiet rooms and shared workspaces, across changes of microphone, and back to earlier settings. Comparisons could examine continued adaptation from personal parameters, reuse of experience from similar settings, and restarts from shared initialization. Opening corrections, subsequent adaptation, and independently collected timing judgments could assess transfer, retention on return, and the additional feedback needed after contextual changes. Recording feedback quantity, content, and distribution, alongside comparisons of initialization, calibration, recent history, and head updating, would help identify which experiences and mechanisms support cumulative adaptation.

\paragraph{Feedback Beyond Taps.}
An important extension is to make timing corrections usable when a touchscreen is inconvenient or inaccessible. Future studies could compare taps with short spoken instructions, such as asking the agent to answer or wait, and with deliberate gestures. The comparison should examine how much effort each form requires, how reliably it is recognized, and whether it can be associated with the response decision the user intended to correct. For example, a spoken request to wait should be distinguished from similar words within the ongoing conversation. This would establish whether alternative controls preserve clear supervision while supporting more natural interaction.
\paragraph{Learning Preference Signals from Multiple Sources.}
Future work could integrate acoustic and prosodic cues, dialogue semantics, interaction states, and correction histories to infer users' preferred response timing. For example, repeated questions and speech overlapping the agent could be interpreted jointly with the preceding exchange and response latency to identify informative episodes. Self-supervised objectives could learn associations between the state at an initiation decision and subsequent user behavior from unlabeled interaction sequences. Adaptive mining could select candidate supervision signals as individual habits and contexts change. Occasional explicit corrections and users' explanations of selected episodes could calibrate those signals to their preferences. Independent timing judgments and later interactions could assess how well the inferred signals match user expectations, their usefulness for online adaptation, and their effects on correction effort.

\paragraph{Feedback Attribution and Conversational Control.}
The reports of desktop interventions during mobile conversations motivate closer study of feedback attribution in settings with several speakers or agents. Future work should establish whose expectations a correction represents and which initiation event it concerns. For example, an optional reason or a brief session-end replay could distinguish an interruption because the user had not finished speaking from one caused by incorrect reply content. Speaker, gaze, or gesture cues could then be evaluated for whether they improve this attribution. Holding the timing mechanism constant while varying dialogue-summary strategies would also help examine how changes in reply content and user expectations influence later corrections.\par Extending learning from initiation to stopping would require a further study of conversational control. When a user begins a new turn during an agent's reply, the system may need to pause, stop, or continue. Evaluations should consider unwanted stops, delays in yielding, recovery after interruption, and the user's ability to override the system. These outcomes would clarify what additional supervision is needed for automatic stopping.

\paragraph{Latency and Privacy in Deployment.}
Deployment studies should distinguish the time at which the system decides to respond from the time at which the user hears the reply. Recording decision, generation, and playback times for the same interaction would help explain whether waiting arises from timing inference or from later processing. Comparisons across devices and network conditions could then assess how those delays affect correction behavior and perceived responsiveness. Local and cloud processing should be compared in terms of response quality, latency, and the information transmitted. For example, keeping speech features on the device may reduce data exposure, but its computational demands and timing performance would need to be measured.\par Further work should examine whether corrective feedback remains useful with full-duplex or end-to-end speech generation. Such an extension would require identifying the initiation event associated with a correction when listening and speaking are managed within one model. Comparisons under matched conversational tasks could assess correction burden and users' ability to intervene, clarifying how the framework's learning and control roles translate to these architectures.

\section{Conclusion}

Speech agents need to adapt to users' differing expectations of response timing during ongoing conversation. Tap-to-Adapt is a framework for interactive speech systems that couples immediate response control with online learning of personalized initiation timing through manual activation and interruption. We developed BPRHT by combining frozen speech representations, class-balanced replay with proximal regularization, and recent correction history for online timing learning. Technical evaluation supported the acoustic-backbone, replay, and history configurations. In the user study, Stage~1 pretrained shared initialization to the prespecified plateau criteria with 20 participants, and Stage~2 evaluated it with 48 new participants. The lower opening correction burden supports shared initialization as a useful starting point. Later rates were similar across conditions. Logs documented system updates from user feedback, and reported speaking and waiting adjustments suggest user accommodation. The pilot corpus supports further evaluation of timing models and learning algorithms.

\section*{Declaration of generative AI and AI-assisted technologies}
ChatGPT and AI-assisted development tools supported code development and execution, reproducible plotting scripts, analysis checks, selected figure components and schematic layout, and drafting and polishing text from the authors' content and outline. Numerical plots use recorded data. The authors reviewed the outputs and take full responsibility for the article, the originality of its ideas and design, and the authenticity of its experiments and data. The Methods section identifies synthetic pilot stimuli.

\bibliographystyle{ACM-Reference-Format}
\bibliography{sample-base}
\end{document}